\documentclass[letterpaper,twocolumn,10pt]{article}
\usepackage{usenix-2020-09}

 \usepackage[T1]{fontenc}
 \usepackage[utf8]{inputenc}

\usepackage{caption}

\usepackage{booktabs}
\usepackage{graphicx}
\usepackage{amsmath}
\usepackage{wasysym}
\usepackage{threeparttable}

\usepackage{upquote}
\usepackage[table]{xcolor} 
\usepackage{multirow}
\usepackage{booktabs}

\usepackage{microtype}
\UseMicrotypeSet[protrusion]{basicmath}
\usepackage{graphicx,grffile}

\usepackage{times} 
\usepackage{tikz}
\usepackage{endnotes,epsfig}
\usepackage{multirow}
\usepackage{xspace}
\usepackage{tabularx}
\usepackage{ragged2e}
\usepackage{booktabs}
\usepackage{paralist}
\usepackage[american]{babel}
\usepackage[shortlabels]{enumitem}
\usepackage{courier,color,wrapfig}
\usepackage{xspace}
\usepackage{balance}
\usepackage{epstopdf}
\usepackage{multirow}
\usepackage{url}
\def\UrlBreaks{\do\/\do-} 
\usepackage{hyperref}
\usepackage{amssymb}
\usepackage{pifont}
\usepackage{subfigure}
\usepackage{tikz}
\usepackage{comment}
\usepackage{mathtools}
\usepackage[normalem]{ulem}
\usepackage{xkeyval}
\usepackage{makecell}
\usepackage{mathrsfs}

\usepackage[most]{tcolorbox}

\usepackage{tikz}
\makeatletter
\let\OldPlane\Plane 
\let\Plane\relax 
\makeatother

\usepackage{tikz-network} 

\makeatletter
\let\Plane\OldPlane 
\makeatother

\usetikzlibrary{arrows.meta}
\usetikzlibrary{automata,positioning}
\usetikzlibrary{calc}
\usetikzlibrary{arrows.meta,arrows,backgrounds,patterns,positioning,shadings, decorations.markings, shapes.multipart, shapes}
\usetikzlibrary{decorations.pathreplacing}

\usepackage{pgfplots}
\usetikzlibrary{patterns,patterns.meta}
\pgfplotsset{compat=newest}
\usepgfplotslibrary{external} 
\usepgfplotslibrary{groupplots} 
\usepgfplotslibrary{statistics}
\usepgfplotslibrary{fillbetween} 
\usepackage{pgfplotstable}

\newif\ifremovecomments

\ifremovecomments
    \usepackage[disable, textsize=small]{todonotes}
\else
    \usepackage[textsize=small]{todonotes}
\fi

\usepackage{ifthen}
\usepackage{etoolbox}

\usepackage{sepfootnotes}
\usepackage[compact]{titlesec}

\usepackage{array}
\newcolumntype{L}[1]{>{\raggedright\let\newline\\\arraybackslash\hspace{0pt}}m{#1}}
\newcolumntype{C}[1]{>{\centering\let\newline\\\arraybackslash\hspace{0pt}}m{#1}}
\newcolumntype{R}[1]{>{\raggedleft\let\newline\\\arraybackslash\hspace{0pt}}m{#1}}
\usepackage{tabularray}
\UseTblrLibrary{booktabs,counter,diagbox,functional,siunitx,varwidth}
\usetikzlibrary{calc}

\usepackage{amsthm}
\usepackage{bbm}
\usepackage[linesnumbered,boxed]{algorithm2e}

\makeatletter
\def\BState{\State\hskip-\ALG@thistlm}
\makeatother

\usepackage[capitalise, nameinlink]{cleveref}
\crefname{algorithm}{Algorithm}{Algorithms}
\Crefname{algorithm}{Algorithm}{Algorithms}
\crefname{section}{\S\!}{\S\S\!}
\Crefname{section}{Section}{Sections}
\crefname{figure}{Figure}{Figures}
\Crefname{figure}{Figure}{Figures}
\crefname{equation}{Equation}{Equations}
\Crefname{equation}{Equation}{Equations}
\crefname{listing}{Listing}{Listings}
\Crefname{listing}{Listing}{Listings}
\crefname{defn}{definition}{definitions}

\makeatletter
\g@addto@macro{\UrlBreaks}{\UrlOrds}
\makeatother

\newcommand\paraspace{\vspace*{0.25ex}}
\providecommand\parab[1]{\paraspace\noindent\textbf{#1}}

\apptocmd\normalsize{%
\abovedisplayskip=5pt
\abovedisplayshortskip=5pt
\belowdisplayskip=5pt
\belowdisplayshortskip=5pt
}{}{}

\pgfplotsset{
    lineerroraxis/.style={
        width=1\columnwidth,
        height=1\columnwidth,
        label style={font=\small},
        legend style={
            font=\small,
            column sep=1pt, 
            row sep=-2pt,
            fill=none,
            draw=none,
            draw opacity=1,
            text opacity=1,
            name=legend,
            at={(0.25,0.95)},
            cells={align=center},
        },
        ymajorgrids=true,
        xmajorgrids=false,
        xtick align=outside,
        xtick pos=bottom,
        x tick style={      
            color=black,      
            line width=0.6pt,
        },
        tick label style={font=\small},
        ytick style={draw=none},
        major tick length=1mm,
        grid style={dashed, lightgray, dash pattern=on 1pt off 1pt},
        set layers,
        ylabel style={yshift=-3pt},
    }
}

\pgfplotsset{
    boxplotstyle/.style={
        width=0.8\columnwidth,
        height=0.8\columnwidth,
        boxplot/draw direction=y,
        ylabel={\large$\frac{\text{OverUtilizedServers}}{\text{Epoch}}$},
        xlabel style={yshift=2pt},
        ylabel style={yshift=-6pt},
        xtick={1,2},
        xticklabels={\small\textsc{Sanjesh}, \small\textsc{Sanjesh}\textsubscript{$M$}},
        grid=both,
        ymajorgrids=true,
        yminorgrids=false,     
        xmajorgrids=false,
        xtick align=outside,
        xtick pos=bottom,
        x tick style={      
            color=black,      
            line width=0.6pt,
        },
        y tick style={      
            color=black,      
            line width=0.6pt,
        },
        ytick pos=left,
        major tick length=1mm,
        minor y tick num=4, 
        minor y tick style={color=gray, line width=0.4pt},
        minor tick length=0.75mm,
        grid style={dashed, lightgray, dash pattern=on 1pt off 1pt},
        boxplot/box extend=0.5,
        ymin=0
    }
}
\newcommand{\sysname}{\textsc{Sanjesh}\xspace}
\newcommand{\simul}{\textsc{Replay}\xspace}
\newcommand{\sa}{\textsc{SA}\xspace}
\newcommand{\genetics}{\textsc{Gen}\xspace}
\newcommand{\hill}{\textsc{HC}\xspace}
\newcommand{\sansim}{\textsc{SanSim}\xspace}
\newcommand{\shuffle}{\textsc{Sanjesh\textsuperscript{Shuffle}}\xspace}
\newcommand{\mone}{$\mathcal{M}_1$\xspace}
\newcommand{\mtwo}{$\mathcal{M}_2$\xspace}

\definecolor{lightgreen}{RGB}{222, 242, 230}
\definecolor{pink}{RGB}{255, 216, 213}
\definecolor{lightblue}{RGB}{216, 237, 240}

\newcommand{\cmark}{\textcolor{green!60!black}{\checkmark}}  
\newcommand{\xmark}{\textcolor{red}{\ding{55}}}  
\newcommand{\pmark}{\textcolor{orange!80!black}{\LEFTcircle}}

\newcommand{\graybox}[1]{%
\begin{tcolorbox}[
  colback=gray!15,
  colframe=gray!40,
  boxrule=0.5pt,
  arc=3pt,
  left=6pt,right=6pt,top=4pt,bottom=4pt,
]
#1
\end{tcolorbox}
}

\newcommand{\rev}[1]{\textcolor{black}{#1}}

\makeatletter
\newcommand{\customlabel}[3]{
   \protected@write \@auxout {}{\string \newlabel {#1}{{#2}{\thepage}{#2}{#1}{}} }
   \hypertarget{#1}{#3}
}

\makeatother

\usepackage{tikz}
\newcommand{\trianglelabel}[3]{%
  \protected@write \@auxout {}{\string \newlabel {#1}{{#2}{\thepage}{#2}{#1}{}} }%
  \hypertarget{#1}{%
    \tikz[baseline=-0.6ex]{
      \node[draw, fill=black, regular polygon, regular polygon sides=3, minimum size=1.5em, inner sep=0pt, text=white] at (0,0) {\small #2};
    }%
  }%
}

\newcommand*\smallcircled[1]{\raisebox{10pt}{\tikz[baseline=(char.center)]{
       \node[inner sep=1pt, draw,blue!70!black, solid, circle, fill=blue!15,text=blue!70!black] (char)
        {\footnotesize #1};}}}
\newcommand*\smallcircledinline[1]{\tikz[baseline=(char.center)]{
       \node[shape=circle,draw,blue!70!black,solid,minimum size=3.3mm, inner sep=0pt, fill=blue!15,text=blue!70!black] (char)
        {\footnotesize #1};}}
\newcommand{\inlinecircle}[1]{\raisebox{3pt}{\smallcircledinline{#1}}}

\newcommand{\cut}[1]{}

\definecolor{set2cyan}{RGB}{114, 182, 161}
\definecolor{set2orange}{RGB}{219, 149, 192}
\definecolor{set2blue}{RGB}{149, 163, 195}
\definecolor{set2purple}{RGB}{219, 149, 192}
\definecolor{set2green}{RGB}{162, 199, 101}
\definecolor{set2yellow}{RGB}{255, 217, 47}
\definecolor{set2brown}{RGB}{229, 196, 148}
\definecolor{set2gray}{RGB}{88, 88, 88}
\definecolor{set2peach}{RGB}{253, 192, 134}
\definecolor{set2green2}{RGB}{127, 201, 127}
\definecolor{tealblue}{RGB}{54, 144, 192}
\definecolor{softcoral}{RGB}{242, 133, 108}
\definecolor{amethystpurple}{RGB}{153, 102, 204}

\definecolor{darkgreen}{RGB}{0, 70, 0}
\definecolor{darkblue}{RGB}{20, 33, 61}
\definecolor{darkred}{rgb}{0.65,0,0}
\definecolor{forestgreen}{rgb}{0.13, 0.65, 0.13}

\definecolor{shapblue}{RGB}{71, 140, 247}
\definecolor{shapred}{RGB}{228, 59, 86}

\usepackage{enumitem}

\begin{document}
\title{A Performance Analyzer for a Public Cloud's ML-Augmented VM Allocator}

\author{%
{\rm Roozbeh Bostandoost}\textsuperscript{$\dagger$},
{\rm Pooria Namyar}\textsuperscript{*},
{\rm Siva Kesava Reddy Kakarla}\textsuperscript{*},
{\rm Ryan Beckett}\textsuperscript{*},\\
{\rm Santiago Segarra}\textsuperscript{\S},
{\rm Eli Cortez}\textsuperscript{\P},
{\rm Ankur Mallick}\textsuperscript{\P},
{\rm Kevin Hsieh}\textsuperscript{*},
{\rm Rodrigo Fonseca}\textsuperscript{*},\\
{\rm Mohammad Hajiesmaili}\textsuperscript{$\dagger$},
{\rm Behnaz Arzani}\textsuperscript{*}\\[0.75ex]
\textsuperscript{$\dagger$}\textit{University of Massachusetts Amherst},
\textsuperscript{*}\textit{Microsoft Research},
\textsuperscript{\S}\textit{Rice University},
\textsuperscript{\P}\textit{Microsoft}
}

\let\textcircled=\pgftextcircled
\maketitle
\begin{abstract}


Cloud operators increasingly deploy multiple ML models in their VM allocation pipelines. In such settings, individually benign predictions can shift and compound, severely degrading performance. In a cloud provider's VM placement pipeline, CPU, memory, and lifetime prediction
models jointly determine server count, live migration frequency, and network utilization; yet no existing approach can systematically stress-test how these models adversely interact. Deterministic adversarial analyzers cannot capture probabilistic ML behavior, so operators miss
failures that arise only from correlated distributional shifts across models

 In \sysname, we formulate a bi-level optimization that captures how the ML models behave statistically and uncovers how they adversely interact. The outer level searches over what predictions the ML models could produce under distributional uncertainty to find adversarial
 conditions; the inner level evaluates how the VM allocator behaves given those predictions. When we applied it to the operator's production traces, \sysname uncovered scenarios that cause $4\times$ worse performance than the operators' evaluator detected.
\end{abstract}


\section{Introduction}
\label{sec:intro}

Operators use machine learning (ML) to improve availability and efficiency~\cite{RC, ishailifetime, dote, danielcaching}. When models underperform, operators naturally focus on individual model accuracy. But in production, ML models compose with other models and heuristics, and the dangerous performance problems are not in any single model. They arise from how individually adequate models interact: correlated prediction shifts that compound through downstream system logic.

\sysname exposes these interaction effects. It reasons about how the predictions of multiple ML models jointly affect downstream system behavior, and finds realistic scenarios where these interactions degrade end-to-end performance.
\rev{We apply \sysname to an ML-augmented virtual machine (VM) allocator~\cite{RC} in a large public cloud that runs at production scale with diverse workloads and rich operational data.} It places incoming 1st-party VM requests on physical servers while minimizing cost and maintaining performance. Users often overestimate their resource needs~\cite{cherrypick}, so the allocator relies on predictions to estimate demand and use servers more efficiently.
The allocator builds features for each VM request and uses a pipeline of ML models to predict resource needs~\cite{RC} (e.g., CPU, memory) and lifetime~\cite{ishailifetime}. It feeds these predictions into a bin-packing heuristic that assigns VMs to servers.



Errors in ML predictions introduce risk. Over-prediction wastes server (and network) capacity, and under-prediction overloads servers—demand exceeds capacity and forces disruptive live migrations that consume network bandwidth and can trigger cascading congestion across the datacenter fabric~\cite{googlelivemigrate, konig2023solver}. These disruptions can cost millions of dollars~\cite{carbonite2015, pingdom2023, nexcess}.

Operators should understand these risks \emph{before} they deploy. They must answer questions like those in \autoref{table:rc_usecases}: which models degrade performance the most, how much does reducing their error help, and how much do their errors increase migration-driven network disruption?


To answer these questions, we need an analyzer that \emph{simultaneously} considers each ML model, its mispredictions, how it interacts with other models, and how it interacts with downstream heuristics.
This is hard: the state space is the Cartesian product of all possible predictions (and mispredictions) for each model and each VM, along with server capacities, which makes it computationally intractable.

\begin{table*}[t]
	\centering
	\small
	\setlength{\tabcolsep}{3pt}
	
	\begin{tabularx}{\textwidth}{
			p{0.28\textwidth}
			*{5}{>{\centering\arraybackslash}X}
			}
		\toprule
		\makecell[l]{\textbf{System (approach)}} &
		\makecell[c]{\textbf{Holistic}\\\textbf{Analysis}} &
		\makecell[c]{\textbf{Performance}\\\textbf{Focus}} &
		\makecell[c]{\textbf{Multi-Model}\\\textbf{Interaction}} &
		\makecell[c]{\textbf{Worst-Case}\\\textbf{Evaluation}} &
		\makecell[c]{\textbf{Feature-Level}\\\textbf{Insight}} \\
		\midrule
		SHAP~\cite{shap} (Feature importance metric) & \xmark & \pmark & \xmark & \xmark & \cmark \\
		MetaOpt~\cite{metaopt} (Performance verifier) & \pmark & \cmark & \xmark & \cmark & \xmark \\
		whiRL~\cite{whirl} (ML safety verifier) & \pmark & \xmark & \xmark & \cmark & \xmark \\
		Pandora~\cite{nushi2018towards} (Empirical analysis) & \cmark & \cmark & \cmark & \xmark & \pmark \\
		\textbf{\sysname (This work)} & \cmark & \cmark & \cmark & \cmark & \cmark \\
		\bottomrule
		\end{tabularx}
	{\footnotesize \pmark = Partially Supported/ Indirectly Addressed}
	\caption{SANJESH provides a more complete understanding of the performance of ML-enabled systems compared to prior work.}
	\label{tab:comparison}
	\end{table*}

Existing solutions, listed in~\autoref{tab:comparison}, fail to address this problem. Empirical approaches~\cite{ishailifetime, ling2025lava, nushi2018towards} replay traces or sample inputs but cannot cover all VM arrival orders and prediction outcomes; in stateful systems, overlooked sequences compound through accumulated state and downstream heuristics, producing far worse behavior than random workloads reveal~(\cref{sec:motivation}).
Formal verification methods~\cite{amir2021towards, amir2023verifying, whirl} check correctness and safety but not \emph{performance} of ML-augmented systems: performance depends on long execution histories and how inputs, predictions, and heuristics interact—not on local invariants~\cite{metaopt, virley}—and probabilistic ML inputs complicate analysis beyond what existing verifiers can handle~\cite{metaopt, virley, mind-the-gap}.
ML explainability tools~\cite{molnar2020interpretable, ale, shap, merrick2020explanation, phillips2021four} focus on individual models and miss multi-model interactions with downstream algorithms (see \cref{sec:relatedwork} for more details on related work).
All three lines of work share a blind spot: they treat models in isolation or assume performance problems stem from specific inputs, not from how the joint prediction distributions of multiple models interact with system logic.


\sysname systematically searches over statistically plausible prediction outcomes to find \emph{realistic} worst-case scenarios—grounded in production distributions, not contrived inputs.
It lets operators pose the distributional queries in \autoref{table:rc_usecases}, including a core open challenge~\cite{fperf, virley}: whether improving one ML model’s accuracy would improve end-to-end performance.

\sysname addresses these queries in two phases: it first finds the model predictions that cause worst-case performance, then maps those predictions back to the feature combinations that produce them.


\begin{table}[!t]
	\centering
	\small
	\SetTblrInner{rowsep=-1pt}
	\begin{tblr}{
			width=\columnwidth,
			colspec={Q[m,wd=0.45cm] X[l,m]},
			hline{1} = {1pt},
			hline{2} = {0.5pt},
			hline{Z} = {1pt},
			row{1} = {font=\bfseries},
		}
		& \textbf{Query/Question} \\
		\customlabel{query:1}{\protect\inlinecircle{1}}{\smallcircled{1}} &
		How much does ML prediction error increase migration risk and server usage compared to an oracle allocator? \\
		\customlabel{query:2}{\protect\inlinecircle{2}}{\smallcircled{2}} &
		Which ML model has the largest impact on allocator performance? \\
		\customlabel{query:3}{\protect\inlinecircle{3}}{\smallcircled{3}} &
		How robust is the system to drifts in VM workload patterns? \\
		\customlabel{query:4}{\protect\inlinecircle{4}}{\smallcircled{4}} &
		Where in the VM feature space does the allocator underperform enough to justify bypassing ML model predictions? \\
		\customlabel{query:5}{\protect\inlinecircle{5}}{\smallcircled{5}} &
		Does an $\mathrm{X}\%$ improvement in an ML model’s accuracy yield enough performance gain to justify its engineering cost? \\
	\end{tblr}
	\caption{The types of queries that \sysname answers for the VM allocator use case (see \cref{sec:sup_queries} for the full set).}
	\label{table:rc_usecases}
\end{table}

\rev{In Phase~1, \sysname builds on MetaOpt~\cite{metaopt}, which generates adversarial inputs via bi-level optimization (a Stackelberg game) to maximize the performance gap between a heuristic and its optimal counterpart. MetaOpt requires deterministic, concrete constraints and cannot express distributional queries over probabilistic ML behavior. \sysname extends this formulation: its outer level searches over what predictions the ML models could produce under distributional uncertainty---enforcing accuracy constraints, correlated error rates, and controlled slack that lets label frequencies deviate by a bounded percentage from observed data---capturing how models interact with downstream system logic~(\cref{sec:phase1}). It also incorporates a scaling technique~(\cref{sec:time-partitioning}) that lets it analyze VM sequences $10\text{--}20\times$ longer than MetaOpt supports.}


In Phase~2, we map problematic predictions back to input features. This is hard because models share features, so per-model attribution misses cross-model interactions. We develop a method based on counterexample guided abstraction refinement (CEGAR)~\cite{clarke2000counterexample} that identifies the minimal feature combinations triggering worst-case predictions, letting operators target mitigations to specific input regions.

\rev{We evaluate \sysname on three days of production traces from a large public cloud VM allocator---millions of VM requests spanning diverse workload patterns. The slack-based constraints explore distributional shifts beyond observed data, so the analysis is not limited to what these traces contain. \sysname surfaces worst-case scenarios with up to $4\times$ worse performance than the operators' existing evaluator detects, and $46\%$ of the high-risk feature combinations it finds appear in the production traces~(\cref{sec:motivation}).}
For each query, \cref{sec:eval} shows how operators can immediately act on \sysname’s output---from guardrail policies to retraining priorities.




\rev{In summary, we make the following contributions:}
\begin{itemize}
	
	\item We identify that the critical risk in ML-augmented infrastructure lies not in individual model errors but in how correlated prediction shifts across models compound through system logic. We formalize this as stress-testing under distributional uncertainty and present \sysname, a two-phase framework that combines a probabilistic Stackelberg formulation (Phase~1) with CEGAR-based feature attribution (Phase~2).
	
	\item We evaluate \sysname on three days of production traces from a large public cloud VM allocator. \sysname finds scenarios with $4\times$ worse performance than the operators' evaluator detects; $46\%$ of the flagged feature combinations appear in production traces~(\cref{sec:eval}).
	
\end{itemize}


We build and evaluate \sysname on a production VM allocator—a complex system with multiple interacting ML models, stateful heuristics, and production-scale workloads—using production traces made available for this research. \rev{Deep evaluation on one production system demonstrates \sysname’s full diagnostic capabilities. The formulation requires only two system-specific inputs: (i) a constraint encoding of the downstream logic and (ii) the ML models’ error distributions derived from held-out data. The Stackelberg formulation, mechanisms, and CEGAR-based attribution are system-agnostic—we chose depth of evaluation over breadth of systems to show the full range of queries \sysname enables. We plan to open-source our tool.}



\section{Motivation: The Blind Spot of Simulation}
\label{sec:motivation}

We build \sysname to analyze a VM allocator~\cite{RC} used in a large production cloud. This allocator relies on two ML models to guide placement decisions. Both are Light Gradient Boosting Machines (LGBMs)~\cite{ke2017lightgbm} and act as binary classifiers:

\begin{itemize}
	\item \textbf{CPU utilization model}: predicts whether a VM has high CPU demand ($\ge 65\%$) or not.
	\item \textbf{Lifetime model}: predicts whether a VM is long-lived (active for $>2$ hours).
\end{itemize}

Each VM request is mapped to features such as VM resource demands, arrival time, and user history. These features are inputs to the ML models. The CPU model estimates how much CPU the VM will use. The lifetime prediction helps the packing heuristic decide which server pool to use. For memory, the allocator uses the requested amount, but we also evaluate a prototype memory model in \cref{sec:eval}.
The allocator feeds these predictions into a \emph{Dynamic Preferred Best-Fit Rule} (DPBFR) heuristic~\cite{ishailifetime} to place a VM onto a server.

DPBFR scores each candidate server by how tightly a VM fits its remaining capacity across all resources (CPU, memory, etc.) and selects from the tightest-fitting bucket. The lifetime prediction controls granularity: long-lived VMs receive finer-grained bucketing to reduce fragmentation (see \cref{sec:vm-allocator} for details on how the VM allocator works).

Mispredictions introduce risk. If the CPU model underpredicts usage, the allocator packs too many VMs onto a server; when they later consume their full resources, demand exceeds capacity and forces live migrations. If the lifetime model mislabels a long-lived VM as short-lived, that VM outlives its neighbors and fragments the server. Operators need to know how much risk these models introduce: how often does over-utilization force migrations, and how many extra servers does it cost compared to an oracle with perfect predictions (\ref{query:1})?

Operators typically rely on trace-driven simulation~\cite{RC, ishailifetime, ling2025lava}: they replay or shuffle historical VM arrivals to estimate the expected number of over-utilized servers. We simulate the allocator on three days of production traces, each containing millions of VM requests (experimental setup in \cref{sec:methodology}).


Replay-based method underestimates risk. DPBFR depends on the \emph{joint} predictions of multiple models across a \emph{sequence} of VMs, and its performance is sensitive to both workload composition and arrival order. Many sequences that are in-distribution for production workloads have not yet appeared in the exact harmful order in the trace. This method explores only a small part of the state space and rarely assembles the combinations of VMs and prediction errors that trigger high migration risk. Formal verification and ML explainability tools also fall short: the former cannot reason about performance over long execution histories with probabilistic inputs, and the latter miss multi-model interactions with downstream heuristics (\autoref{tab:comparison}). 
To expose this blind spot and show that the risks are real, we compare four validation approaches in \autoref{fig:motivation}:

\begin{figure}
	\centering
	\begin{tikzpicture}
		\definecolor{cpuBlue}{RGB}{33,98,173}
		\definecolor{memGold}{RGB}{246,181,0}
		\begin{axis}[
			ybar=0pt, 
			bar width=20pt,
			width=0.98\columnwidth,
			height=0.5\columnwidth,
			ymin=0,
			ymax=4.5,
			enlarge x limits=0.20,
			ymajorgrids,
			grid style={gray!60, densely dashed},
			tick style={black},
			xtick pos=bottom,
			symbolic x coords={SANJESH,SHUFFLE,SANSIM,SIMUL},
			xtick={SANJESH,SHUFFLE,SANSIM,SIMUL},
			xticklabels={\sysname,\shuffle,\sansim,\simul},
			xticklabel style={align=center, font=\footnotesize},
			ylabel={\small Risk of Migration},
			ytick={1,2,3,4},
			yticklabels={{1x},{2x},{3x},{4x}},
			ytick pos=left,
		]
			\addplot+[bar shift=0pt]
			coordinates {(SANJESH,4)};
			
			\addplot+[bar shift=0pt]
			coordinates {(SHUFFLE,3.5)};
			
			\addplot+[bar shift=0pt]
			coordinates {(SANSIM,2.1)};
			
			\addplot+[bar shift=0pt]
			coordinates {(SIMUL,1)};
			
		\end{axis}
	\end{tikzpicture}
	\caption{Trace replay (\simul) underestimates migration risk (performance degradation). \sysname finds scenarios with up to $4\times$ higher risk under realistic workload constraints. \shuffle randomizes arrival order and reduces risk, but it remains substantial. \sansim uses the subset of \sysname VMs present in the trace and fills the rest from the trace, yielding $\sim 2\times$ higher risk than \simul. The y-axis shows median migration risk normalized to \simul.}
	\label{fig:motivation}
\end{figure}


\begin{enumerate}
    \item \textbf{\simul:} Standard trace replay.
	\item \textbf{\sysname:} Finds \emph{in-distribution} scenarios, consistent with VM arrivals in production trace, that lead to worst-case performance (experimental setup in~\cref{sec:methodology}).
	\item \textbf{\shuffle:} Uses the same VMs as \sysname but randomizes their arrival order.
	\item \textbf{\sansim:} Keeps only the VMs from the \sysname sequence that appear in the production trace and replaces the rest with random samples from the trace. It also randomizes the order. 
\end{enumerate}


\sysname uncovers scenarios where migration risk due to server overutilization is up to $4\times$ higher than what \simul reports. These scenarios are not outliers. About $46\%$ of the adversarial VM arrival sequences \sysname generates have close matches in the production trace, differing only in small shifts in arrival time ($\le 20$ minutes).

\graybox{\textbf{Observation:} The individual VMs and their features already exist in practice. \simul misses the specific combinations and arrival orders that trigger failures. Prior work on large-scale networked systems has shown that rare but harmful input combinations inevitably occur at scale~\cite{metaopt, raha, groot}; the question is whether operators detect them before they cause costly incidents.}




Two variants of the \sysname sequence sharpen the analysis. Does \sysname rely on a specific arrival order? \shuffle shows it does not: randomizing the order preserves most of the risk ($\sim 3.5\times$, \autoref{fig:motivation}). The allocator struggles with \emph{structurally challenging workload compositions}—sets of valid VMs that are hard to place efficiently, regardless of order. Arrival order amplifies the effect (from $3.5\times$ to $4\times$), but composition drives it.

\graybox{\textbf{Observation:} High migration risk does not depend on a single arrival order. It comes from specific combinations of valid VM requests that stress the allocator. Arrival order only changes how severe the impact is.}




The \sansim experiment further confirms that replay-based methods miss real risk. The trace replay underestimates risk by $\sim 2\times$ compared to \sansim, even though both draw only from trace-backed VMs. The difference is \emph{selection}: simulation samples arrivals without guidance, whereas \sansim seeds the sequence with the problematic VMs that \sysname identifies and fills the rest from the trace.

The remaining gap between \sysname and \sansim comes from additional in-distribution VMs that further stress the allocator. Such reinforcing arrivals can occur during bursts, workload shifts, or correlated submissions.

\graybox{\textbf{Observation:} \simul misses high-risk scenarios because it does not search for problematic subsets of requests. \sysname provides the guidance that lets \sansim expose risks already present in the trace but hidden.}


Together, these results show that replay-based method systematically underestimates operational risk by failing to explore harmful but realistic interactions between VM arrivals, ML predictions, and system behavior. Operators need a tool like \sysname to surface these vulnerabilities before they cause costly production incidents. We compare \sysname to more sophisticated baselines in~\cref{sec:eval}.

\section{\sysname Overview}
\label{sec:sanjesh}

\sysname identifies VM workloads and ML model predictions that cause worst-case system performance compared to an oracle with access to ground-truth labels. 
\sysname operates in two phases. First, it identifies prediction patterns that maximize performance degradation. Second, it maps these predictions to the VM features that produce them.

\rev{Phase~1 builds on MetaOpt~\cite{metaopt}, which generates adversarial inputs for heuristic systems via Stackelberg games. MetaOpt assumes deterministic inputs and analyzes a single heuristic; \sysname extends the formulation to search over probabilistic prediction distributions of multiple interacting ML models.} \sysname formulates this phase as a leader–follower (Stackelberg) game. The leader selects ML model prediction outcomes (and corresponding ground-truth labels) to maximize system degradation. The followers execute the VM allocator under two settings: (i) the production system that uses ML predictions, and (ii) an oracle that uses ground truth. The difference between these outcomes captures the impact of prediction errors.

We express this interaction as a bi-level optimization:

\begin{equation}
	\begin{aligned}
		&\text{Outer problem: } 
		\max_{I, I' \in \mathscr{I}} \; H'(I') - H(I) \\
		&\text{s.t. } 
		\text{Constraints}(I, I', \text{Query}, \text{ML Models}, \text{Trace})
	\end{aligned}
	\label{eq:bi-level}
\end{equation}
\vspace{-0.1cm}
\begin{equation*}
	\begin{aligned}
		\text{Inner Problem 1: } & H'(I') = \text{ML-based system}(I') \\
		\text{Inner Problem 2: } & H(I) = \text{Oracle}(I)
	\end{aligned}
\end{equation*}

\begin{figure*}[t]
	\centering
	\includegraphics[width=\linewidth]
	{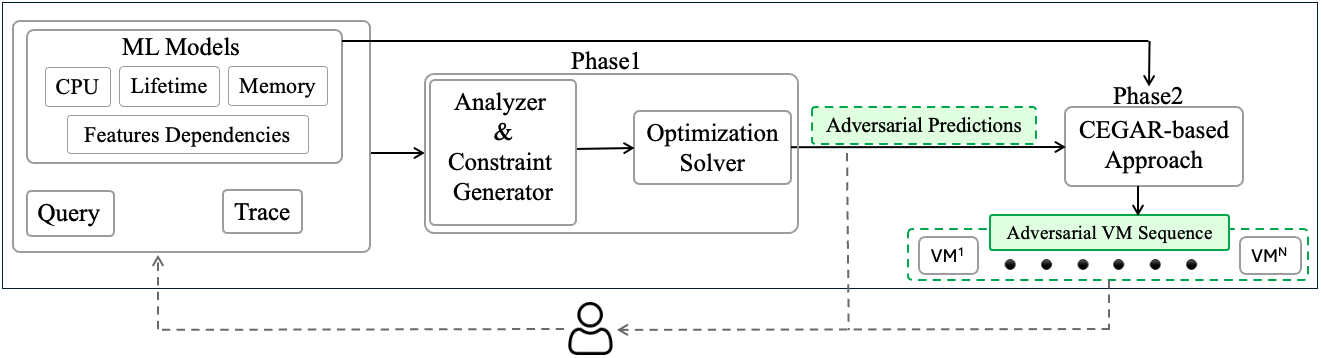}
	\caption{\sysname overview. Users provide the ML models, their feature dependencies, and production trace, along with the specific queries for \sysname to answer (\autoref{table:rc_usecases}). In Phase~1, the analyzer and constraint generator converts these inputs into optimization constraints and formulates the problem. The optimization solver computes adversarial predictions of the models and ground-truth labels that lead the VM allocator to underperform. In Phase~2, a CEGAR-based approach derives VM feature sequences that produce these predictions. \sysname outputs the resulting VM sequences and predictions to the user.}
	\label{fig:sanjesh_overview}
\end{figure*}

Here, $I’$ represents the ML model predictions for a VM sequence, and $I$ represents the corresponding ground-truth labels. The functions $H’$ and $H$ measure system-level outcomes, such as migration risk (e.g., number of over-utilized servers) or server usage under the ML-based system and the oracle.

These constraints restrict the search to plausible workloads by enforcing model accuracy, correlated errors, and workload distributions derived from production traces.

Crucially, the outer problem searches over the joint prediction outcomes of all models simultaneously. A CPU underprediction and a lifetime overprediction may each be individually plausible, but when they co-occur on the same VM, the allocator places a resource-hungry, long-lived VM on an already-strained server---compounding fragmentation and overload. The joint search over $I$ and $I'$ exposes these effects.

The bi-level formulation is not specific to the VM allocation problem; we discuss how it extends to other systems in \cref{sec:discussion}.

\autoref{fig:sanjesh_overview} shows the system design. \sysname takes as input (i) the operator’s query (\autoref{table:rc_usecases}), (ii) the ML models, and (iii) sample production data. In Phase~1, the analyzer and constraint generator convert these inputs into probabilistic constraints (\cref{sec:phase1}) and encode the allocator logic and performance metrics (\cref{sec:dpbfr}). It then formulates the problem as a mixed-integer program (\autoref{eq:bi-level}) and passes it to the optimization solver~\cite{gurobi}, which outputs prediction patterns and ground-truth labels that expose worst-case behavior.

In Phase~2, \sysname maps these predictions to the VM features that produce them. We encode the CPU and lifetime LGBMs in an SMT solver~\cite{z3} to find feature assignments that yield the target predictions for each VM. \rev{We use a CEGAR-based approach (\cref{sec:phase2}) to scale this step, since models share features and the tree ensembles are large, which makes direct inversion expensive.}





\section{Phase 1: Finding Worst-Case Predictions}
\label{sec:phase1}

Phase~1 solves the bi-level optimization in \autoref{eq:bi-level} to identify prediction sequences $(I, I')$ that maximize the performance gap $H'(I') - H(I)$.
The structure of the inner problems depends on the query. For queries~\ref{query:1}--\ref{query:4}, $H(I)$ represents the VM allocator operating with ground-truth labels (the oracle), while $H'(I')$ corresponds to the allocator using ML predictions. For query~\ref{query:5}, $H(I)$ models the allocator with the current CPU model, and $H'(I')$ uses a hypothetical improved model. In all cases, the outer problem maximizes the gap between these two systems.



The analyzer translates queries (\autoref{table:rc_usecases}), ML model behavior, and workload characteristics into optimization constraints. We develop five mechanisms ($M_1$--$M_5$) to address three challenges: encoding realistic ML models ($M_1$, $M_2$), modeling uncertainty ($M_3$, $M_4$), and scaling to long VM sequences ($M_5$).

\subsection{Challenge: Encoding Realistic ML Models}
\label{sec:realistic}

Queries~\ref{query:1}--\ref{query:5} require the search to stay within \emph{realistic} production patterns. $M_1$ and $M_2$ enforce this.

\parab{Mechanism $M_1$: Extracting typical behavior.}
$M_1$ mines production traces to build a statistical profile of the system: how often each model predicts correctly, how errors correlate across models (e.g., the rate at which the CPU and lifetime models both err on the same VM), the distribution of ground-truth labels, and the mix of VM types and arrival rates over time. This defines what ``realistic'' means for the optimizer.

\parab{Mechanism $M_2$: Enforcing realistic constraints.}
Consider a CPU model that is correct on $70\%$ of VMs. In a sequence of $N=10$ VMs, exactly $7$ correct predictions is only one possibility; the true count is random. $M_2$ uses the Central Limit Theorem~\cite{kwak2017central} and Chebyshev's inequality~\cite{chebychev} to compute a range $[6, 8]$ that covers $99.99\%$ of plausible outcomes (constraint~B in \autoref{fig:mechanisms_examples}). It also determines the minimum sequence length $N$ (Constraint~A in \autoref{fig:mechanisms_examples}) for which these bounds are tight enough to be useful (details in \cref{sec:probsfromsamples}).

More generally, $M_2$ converts every distribution from $M_1$---model accuracy, label frequencies, correlated error rates---into count-bounded constraints on $I$ and $I'$ in \autoref{eq:bi-level}. The optimizer may place the counts anywhere within the allowed ranges, but it cannot fabricate scenarios where every model fails at once.

\begin{figure}[t]
	\centering
	\resizebox{\columnwidth}{!}{
\definecolor{badgeA}{RGB}{210,235,190} 
\definecolor{badgeB}{RGB}{240,210,235} 
\definecolor{badgeC}{RGB}{205,230,245} 
\definecolor{badgeD}{RGB}{255,215,195} 

\tikzset{
  panel/.style = {rounded corners=3pt, draw=black!60, very thick, fill=white},
  subpanel/.style = {rounded corners=4pt, draw=black!40, thick, fill=white},
  title/.style = {font=\bfseries},
  badge/.style = {
      circle, minimum size=11pt, inner sep=0pt,
      font=\small, draw=black!40
  }
}

\begin{tikzpicture}[trim left]
\footnotesize

\node[panel, minimum width=4.5cm, minimum height=2.75cm, anchor=north west]
  (L) at (0,0) {};

\node[title, anchor=south west] at ($(L.north west)+(0,0)$) {Constraint (Mechanism)};

\node[badge, fill=badgeA] (A) at ($(L.west)+(0.3,1.1)$) {A};
\node[anchor=west] at ($(A.east)+(-0.05,0)$)
  {$N = 10$};
\node[anchor=west] at ($(A.east)+(3.0,0)$)
  {$M_1, M_2$};
  
\node[badge, fill=badgeB] (B) at ($(A.south)+(0,-0.5)$) {B};
\node[anchor=west] at ($(B.east)+(-0.05,0)$)
  {$6 \le \sum_{i=1}^{N} x_i \le 8$};
\node[anchor=west] at ($(B.east)+(3.0,0)$)
  {$M_1, M_2$};
  
\node[badge, fill=badgeC] (C) at ($(B.south)+(0,-0.5)$) {C};
\node[anchor=west] at ($(C.east)+(-0.05,0)$)
  {$3 \le \sum_{i=1}^{N} y_i \le 9$};
\node[anchor=west] at ($(C.east)+(3.0,0)$)
  {$M_1, M_4$};

\node[badge, fill=badgeD] (D) at ($(C.south)+(0,-0.5)$) {D};
\node[anchor=west, align=left] at ($(D.east)+(-0.05,0)$)
  {$x_j = 1 \wedge x'_j = 1,\ \forall j \in \mathcal{R}_6$};
\node[anchor=west] at ($(D.east)+(3.0,0)$)
  {$M_1, M_5$};

\node[panel, minimum width=3.95cm, minimum height=2.9cm, anchor=north west]
  (R) at ($(L.north east)+(0.3,0)$) {};

\node[title, anchor=south west] at ($(R.north west)+(0,0)$) {Use case};

\node[subpanel, minimum width=3.65cm, minimum height=0.5cm, anchor=north west]
  (R1) at ($(R.north west)+(0.15,-0.1)$) {};
\node[anchor=west, font=\footnotesize, align=left] at ($(R1.north west)+(0.05,-0.25)$)
  {CPU model accuracy};
\node[badge, fill=badgeA, anchor=west] at ($(R1.east)+(-1.0,0)$) {A};
\node[badge, fill=badgeB, anchor=west] at ($(R1.east)+(-0.5,0)$) {B};

\node[subpanel, minimum width=3.65cm, minimum height=1cm, anchor=north west]
  (R2) at ($(R.north west)+(0.15,-0.7)$) {};
\node[anchor=west, font=\footnotesize, align=left] at ($(R2.north west)+(0.05,-0.5)$)
  {Drift in CPU model \\ data (Label 0)};
\node[badge, fill=badgeA, anchor=west] at ($(R2.east)+(-0.5,0.25)$) {A};
\node[badge, fill=badgeC, anchor=west] at ($(R2.east)+(-0.5,-0.25)$) {C};

\node[subpanel, minimum width=3.65cm, minimum height=1cm, anchor=north west]
  (R3) at ($(R.north west)+(0.15,-1.8)$) {};
\node[anchor=west, font=\footnotesize, align=left] at ($(R3.north west)+(-0.05,-0.5)$)
  {Current and hypothetical \\ CPU models joint accuracy};
\node[badge, fill=badgeA, anchor=west] at ($(R3.east)+(-0.5,0.25)$) {A};
\node[badge, fill=badgeD, anchor=west] at ($(R3.east)+(-0.5,-0.25)$) {D};

\draw[dashed, -{Stealth[length=2mm]}, thick]
  ($(L.east)+(0.0,-0.07)$) -- ($(R.west)+(0.0,-0)$);

\end{tikzpicture}}
	\caption{Constraint encoding examples. $N = 10$ VMs. $x_i = 1$ (binary) when the CPU model's prediction matches the ground truth for $\text{VM}_i$ ($x'$ for the hypothetical model). $y_i = 1$ (binary) when the ground-truth label for CPU model is~0. $\mathcal{R}_6$ is a random sample of 6 indices from $\{1,\ldots,10\}$.}
	\label{fig:mechanisms_examples}
\end{figure}

\subsection{Challenge: Modeling Uncertainty}
\label{sec:query_encoding}


Some queries require reasoning about conditions that differ from current production behavior. $M_3$ and $M_4$ extend the formulation to handle two forms of uncertainty: workload drift and hypothetical model improvements.


\parab{Mechanism $M_3$: Data drift.}
Query~\ref{query:3} asks: how much can the workload shift before the system breaks? $M_3$ answers this by loosening $M_2$'s constraints---widening the allowed ranges so the optimizer can explore post-drift workloads.

For example, suppose that label~0 of the CPU model has a frequency of $60\%$. For $N=10$ VMs, $M_2$ enforces $[5,7]$ VMs with that label. If the operator specifies $20\%$ drift, $M_3$ widens this by $\pm2$ VMs to $[3,9]$---allowing the optimizer to explore workloads where the label mix has shifted substantially (Constraint~C in \autoref{fig:mechanisms_examples}).

\parab{Mechanism $M_4$: Hypothetical model.}
Query~\ref{query:5} asks: would improving a model's accuracy by $X\%$ justify its engineering cost---without training the improved model? The difficulty is that the current model's accuracy, the hypothetical model's accuracy, and how often the two agree are not independent. For instance, if the current model is $70\%$ accurate and the hypothetical is $80\%$ accurate, they cannot agree on fewer than $50\%$ of VMs (since both are mostly correct). Setting these rates independently produces contradictory constraints.

$M_4$ resolves this with a small linear program that finds a consistent \emph{joint distribution} over prediction outcomes---a table whose entries are the fraction of VMs where both models are correct, only one is correct, or both are wrong (details in \cref{sec:flow_model_hypothetical}). If no consistent table exists, the operator's targets are contradictory and $M_4$ reports this. Otherwise, $M_4$ feeds each cell of the table as a count-bounded constraint (via $M_2$) into the optimization. Constraint~D in \autoref{fig:mechanisms_examples} shows one such cell. \autoref{table:query_mechanisms} summarizes which mechanisms each query uses.

\begin{table}[t]
	\centering
	\footnotesize
	\SetTblrInner{rowsep=1pt, colsep=3pt}
	\begin{tblr}{
			width=\columnwidth,
			colspec={Q[m,wd=0.45cm] X[l,m] Q[l,m,wd=2.0cm]},
			hline{1} = {1pt},
			hline{2} = {0.5pt},
			hline{Z} = {1pt},
			row{1} = {font=\bfseries},
		}
		& \textbf{$(H', H)$ pair} & \textbf{Active mechanisms} \\
		\ref{query:1} & ML allocator vs.\ oracle & $M_1$, $M_2$, $M_5$ \\
		\ref{query:2} & ML allocator (one model) vs.\ oracle & $M_1$, $M_2$, $M_5$ \\
		\ref{query:3} & ML allocator vs.\ oracle under drift & $M_1$, $M_2$, $M_3$, $M_5$ \\
		\ref{query:4} & ML allocator vs.\ oracle + risk surface & $M_1$, $M_2$, $M_5$ \\
		\ref{query:5} & Hypothetical allocator vs.\ current allocator & $M_1$, $M_2$, $M_4$, $M_5$ \\
	\end{tblr}
	\caption{$M_1$, $M_2$, $M_5$ are active in every query: all analyses require realistic error distributions, finite-sequence bounds, and time-partitioning scaling. $M_3$, $M_4$ are activated only by the queries that specifically target drift and hypothetical model behavior. The risk surface for query~\ref{query:4} is computed in a post-processing step (\cref{sec:risk_surface_main}).}
	\label{table:query_mechanisms}
\end{table}

\subsection{Challenge: Scaling to Long VM Sequences}
\label{sec:time-partitioning}





The constraints that mechanisms enforce require sequences of at least $N \sim 100$--$200$ VMs for the bounds to be statistically meaningful. But, the optimization solver~\cite{gurobi} can handle bin-packing instances of only $10$--$20$ VMs within a few hours; at $N=200$, it fails to return a solution even after 24 hours.

\parab{Mechanism $M_5$: Time-based partitioning.}
$M_5$ addresses this by splitting the $N$ VMs into $\mathcal{K}$ groups based on arrival time. It solves the bi-level optimization on each group and carries the server state forward across groups. Smaller groups solve faster but miss long-range interactions; larger groups capture more but increase runtime. Because $M_5$ carries server state forward, interactions that depend on cumulative load---such as a server gradually overloaded by VMs from multiple groups---propagate across partitions even when individual groups are small.

$M_5$ manages this trade-off with warm-starting~\cite{gurobi}: it first solves with small groups, then uses that solution to initialize the same problem with larger groups.
The solver reaches a quality solution faster from a warm start.

$M_5$ also distributes global constraints across groups. For example, for $N=10$ and $\mathcal{K}=2$ (two groups of 5 VMs), $M_2$ requires $6$--$8$ correct CPU predictions overall. $M_5$ enforces at least $6/2=3$ correct predictions per group and ensures that the total across both groups does not exceed 8.

\section{Phase 2: Finding Feature Vectors}
\label{sec:phase2}

Phase~1 outputs a prediction sequence: for each of the $N$ VMs, the ML model predictions and the corresponding ground-truth labels.
Phase~2 must find feature vectors $\textbf{f}$ that, when passed to the deployed LGBMs, produce those exact predictions.

\subsection{Why This Is Hard}
Two straightforward approaches fail. Searching the training data for matching examples rarely satisfies all models simultaneously because models share input features. Encoding the LGBMs directly as mixed-integer constraints (\cref{sec:LGBM}) works for small models but grows exponentially with tree depth---infeasible for the deep production LGBMs the allocator uses.

The difficulty stems from two properties:

\parab{Shared input features.}
The CPU, lifetime (and optionally memory) models share features such as VM size (resources demand, e.g., CPU cores and memory) and arrival time. Phase~2 must find a single feature vector $\textbf{f}$ that simultaneously produces the desired predictions for all models. This creates a joint constraint that cannot be decomposed into independent per-model searches.

\parab{LGBM structure.}
Each LGBM computes a probability distribution over classes: it sums the leaf scores of all trees associated with each class, applies a softmax to obtain per-class probabilities, and predicts the class with the highest probability.
A single tree's leaf value is meaningless without the others, since the final class probabilities depend on the aggregate scores across all trees.
This coupling rules out per-tree analysis techniques designed for random forests, where each tree votes independently~\cite{ceagr1, devos2021versatile}.
We cannot fix a target for one tree and analyze it in isolation.
\subsection{Solution: Scalable Joint Feature Inference}
\label{sec:CEGAR}

The key insight is to start with a coarse approximation of each model and refine only where it fails---analogous to binary search on the model structure, where each iteration halves the remaining uncertainty.

We encode all LGBMs together in the Z3 SMT solver~\cite{z3}.
Each tree is represented as a set of threshold tests on input features that lead to a leaf score.
The solver finds a single feature assignment that drives all trees of all models to scores that, after aggregation and softmax, yield the target class as the highest-probability prediction for each model (details in \cref{sec:lgbm_smt}).

\rev{Directly solving this formulation is expensive for large LGBMs with deep trees. We scale it using a novel CEGAR-based approach~\cite{clarke2000counterexample}, which incrementally refines the model to focus only on relevant parts of the search space.} The approach proceeds in three steps:

\parab{Step 1: Abstract all trees.}
We prune every tree in each model to a fixed shallow depth. For non-target classes, we replace deeper leaves with the minimum possible leaf value across pruned nodes; for the target class, we use the maximum. This makes the abstraction conservative with respect to reaching the target predictions.
The key property is that if the SMT query is infeasible on the abstracted trees, it is also infeasible on the full model. In that case, no feature vector can produce the target predictions (\autoref{fig:cegar_trees}).

\parab{Step 2: Solve for features.}
The solver searches for a single feature vector that leads to target predictions across all models. If the query is infeasible under the current abstraction, we conclude the target predictions are unreachable and discard them from Phase~1.

\begin{figure*}[t]
	\centering
	\resizebox{\textwidth}{!}{%
		\begin{tikzpicture}[
			level distance=0.75cm,
			edge from parent/.style={draw,-latex},
			level 1/.style={sibling distance=2.5cm},
			level 2/.style={sibling distance=1.5cm},
			level 3/.style={sibling distance=1.5cm},
			every node/.style={draw, minimum size=3mm, font=\small},
			feature/.style={circle, inner sep=0.06cm, fill=gray!7},
			leaf/.style={rectangle, fill=white},
			]
			\small
			\begin{scope}[local bounding box=T1, xshift=0cm]
				\node[feature] (root) {$f_1$}
				child {node[feature] (f2) {$f_2$}
					child {node[leaf] (l1) {0.9}}
					child {node[feature] (f3) {$f_3$}
						child {node[leaf] (l2) {1.3}}
						child {node[leaf, fill=blue!20] (l3) {0.5}}
					}
				}
				child {node[leaf, fill=blue!20] (l4) {0.6}};
				\node[draw=none] at (-2, 0) {\textsc{Class} 1};
			\end{scope}
			\begin{scope}[local bounding box=T2, xshift=5.8cm]
				\node[feature] {$f_1$}
				child {node[feature] (f23) {$f_3$}
					child {node[feature] (f22) {$f_2$}
						child {node[leaf, fill=forestgreen!50] (l21) {1.1}}
						child {node[leaf] (l22) {0.4}}
					}
					child {node[leaf] (l23) {0.2}}
				}
				child {node[feature] (f222) {$f_2$}
					child {node[leaf] (l24) {0.8}}
					child {node[leaf, fill=forestgreen!50] (l25) {1.0}}
				};
				\node[draw=none] at (-2, 0) {\textbf{\textsc{Class} 2 (target)}};
			\end{scope}
			\begin{scope}[local bounding box=T3, xshift=12cm]
				\node[feature] {$f_2$}
				child {node[feature] (f31) {$f_1$}
					child {node[leaf, fill=blue!20] (l31) {0.3}}
					child {node[feature] (f33) {$f_3$}
						child {node[leaf] (l32) {1.4}}
						child {node[leaf] (l33) {0.7}}
					}
				}
				child {node[leaf, fill=blue!20] {0.5}};
				\node[draw=none] at (-2, 0) {\textsc{Class} 3};
			\end{scope}
			\begin{pgfonlayer}{background}
				\node[draw, rounded corners, fit=(T1), inner sep=8pt] {};
				\node[draw=blue!80, dashed, thick, rounded corners,
				fit=(f2)(f3)(l1)(l2)(l3), inner sep=3pt] {};
				\node[draw, rounded corners, fit=(T2), inner sep=8pt] {};
				\node[draw=forestgreen, dotted, thick, rounded corners,
				fit=(f23)(f22)(l21)(l22)(l23), inner sep=3pt] {};
				\node[draw=forestgreen, dotted, thick, rounded corners,
				fit=(f222)(l24)(l25), inner sep=3pt] {};
				\node[draw, rounded corners, fit=(T3), inner sep=8pt] {};
				\node[draw=blue!80, dashed, thick, rounded corners,
				fit=(f31)(f33)(l31)(l32)(l33), inner sep=3pt] {};
			\end{pgfonlayer}
		\end{tikzpicture}
	}
	\caption{CEGAR-based approach applied to a three-class LGBM with one tree per class and features $f_1, f_2, f_3$.
		Goal: find a feature vector that makes the model predict \textsc{Class}~2.
		We prune each tree to depth~1.
		Pruned regions of Class~1 and Class~3 trees use minimum leaf values (blue, dashed) to underestimate their scores; pruned regions of the Class~2 tree use maximum values (green, dotted) to overestimate its score.
		If the SMT query is infeasible under this abstraction, \textsc{Class}~2 is provably unreachable in the full model.
		If a candidate is found, we verify it on the full trees and expand the abstraction where it fails.
		In the multi-model case (CPU model + lifetime model), the same SMT query encodes both models with shared feature variables, so constraints added while refining one model immediately propagate to the other.}
	\label{fig:cegar_trees}
\end{figure*}

\parab{Step 3: Verify and refine.}
If the solver returns a candidate feature vector, we evaluate it on the full-depth trees of all models. If every model produces its target prediction, Phase~2 is complete for that VM. If any model disagrees, the mismatch finds the tree and branch where the abstraction is too coarse. We refine the it along that branch and repeat from Step~1.

The procedure terminates because each refinement strictly increases the abstraction depth; it either finds a valid feature assignment or proves none exists. By refining only the branches that cause mismatches, it does not explore the full model. We compare it to direct SMT solving in \cref{sec:phase2_perf} and describe how to jointly enforce ground-truth labels in \cref{sec:ground-truth-cegar}.

\section{Post-Analysis: Computing the Risk Surface}
\label{sec:risk_surface_main}

After Phase~2 identifies feature vectors that cause worst-case behavior, query~\ref{query:4} asks: \emph{where} in the feature space does the allocator underperform?

Each adversarial feature vector from Phase~2 lands in a leaf region of the LGBM trees---a hyperrectangle in feature space where any input produces the same prediction. Many distinct feature vectors map to the same region. We aggregate these regions across the $N$-VM sequence to construct a \emph{risk surface} (\cref{sec:risk_surface}): the union of feature-space regions where high-risk behavior concentrates.

Operators use the risk surface to design guard conditions. For example, if high-risk scenarios cluster within a specific range of a feature, the system can bypass ML predictions for VMs in that range and allocate conservatively (e.g., based on requested values).

\section{Evaluation}
\label{sec:eval}


\sysname is the first system that analyzes how multiple ML models \rev{jointly affect end-to-end VM allocator performance}. We show: (1) how \sysname evaluates performance under realistic workloads and answers operator-driven queries, and (2) how it outperforms the baselines.


\subsection{Methodology}
\label{sec:methodology}

We use the Gurobi optimizer~\cite{metaopt} to solve the Stackelberg formulation in \autoref{eq:bi-level} for Phase~1, and Z3~\cite{z3} as the SMT solver for Phase~2. \rev{Mechanism $M_2$ determines that at least sequences of $N=200$ VMs are required for the constraints in Phase~1 to be statistically meaningful based on production traces and model behavior; we use sequences of this length in our analysis.}
We divide time into 10-minute epochs and analyze the system state at each epoch over a full day (144 epochs). We repeat each experiment until the results reach statistical significance (Mann-Whitney U test, $\alpha = 0.1$).
\sysname analyzes VM sequences using time-partitioning technique (\cref{sec:time-partitioning}): it processes 5 VMs per partition and applies a 30-minute time limit to each partition.

\parab{Datasets.} We use data from three weekdays in July 2024 and the corresponding CPU and lifetime models from a VM allocator in a large public cloud. The slack-based constraints in Phase~1 let \sysname explore prediction distributions beyond what these three days contain, so the analysis stress-tests behavior within a controlled neighborhood of the observed data rather than being limited to the literal traces. We also evaluate a prototype memory model under development by the operators. This model is a multi-class classifier that predicts the fraction of requested memory a VM will use. We train the models on data from the first day and use the remaining days to estimate the distributions to enforce in each query.


\parab{Metric.} We use the \emph{risk of migration} to measure allocator performance: the average number of over-utilized servers per epoch. We normalize all results to \sysname’s median, which provides a stable reference point for comparing across experiments (the absolute number of over-utilized servers varies with trace characteristics).

\parab{Benchmarks.} 
To our knowledge, no prior work can perform end-to-end analysis of multi-model ML-based systems and answer all queries in~\autoref{table:rc_usecases}. Operators and researchers typically rely on simulation-based approaches or analyze each model in isolation. We compare \sysname against these baselines under the same runtime budget. Each baseline evaluates the same number of VM sequences as \sysname and runs them through the allocator with ML predictions and with perfect predictions (oracle) to measure the risk of migration:

\parab{(1) Trace Replay (\simul):} We randomly sample VM sequences from the trace and report the risk of migration.

\parab{(2) Hill Climbing (\hill):} We implement a greedy search baseline that explores VM sequences through local modifications. Starting from a sampled sequence from the trace, each iteration applies either (i) composition changes, which replace one or two VMs while preserving realism constraints, or (ii) reordering changes, such as swaps, insertions, or block moves. We only accept sequences with higher migration risk.

\parab{(3) Simulated Annealing (\sa):} We use the same search space and neighborhood as \hill, but replace the acceptance rule with simulated annealing. In addition to improvements, \sa occasionally accepts worse candidates under a gradually decreasing acceptance schedule, allowing it to escape local optima and explore a broader set of high-risk sequences.

\parab{(4) Genetic Algorithm (\genetics):} We implement a population-based search baseline that evolves VM sequences over time. Each chromosome represents an ordered sequence of VMs. We initialize a population with feasible sequences sampled from the trace and evaluate each using the same risk metric.

At each iteration, we select parent sequences using tournament selection and generate new candidates through crossover and mutation. The crossover combines two parent sequences position-wise, while mutation applies either (i) composition changes, which replace a small number of VMs while preserving realism constraints, or (ii) reordering changes, such as swaps, insertions, or block moves. We use elitism to retain high-quality sequences and replace weaker ones with improved candidates.

\begin{figure*}[t]
	\centering
	\begin{tikzpicture}
		\begin{groupplot}[
			group style={group size=2 by 1, horizontal sep=1.5cm},
			height=0.18\textwidth,
			scale only axis,
			ymin=0, ymax=1.5,
			ylabel={Risk of Migration},
			ytick={0,0.25,0.5,0.75,1,1.25,1.5},
			yticklabel style={font=\small},
			xticklabel style={font=\small},
			]
			
			
			\nextgroupplot[
			width=0.36\textwidth,
			xlabel={\small Elapsed Run Time (hours)},
			xmin=0, xmax=19,
			ymin=0, ymax=1.5,
			legend style={
				at={(0.5,1.05)},
				anchor=south,
				legend columns=-1,
				font=\footnotesize,
				draw=none,
				fill=none,
			},
			xlabel style={yshift=2pt},
			ylabel style={yshift=-6pt, xshift=-6pt},
			grid=both,
			ymajorgrids=true,
			yminorgrids=false,
            xtick={0,5,10,15,19},
			xtick align=outside,
			xtick pos=bottom,
			x tick style={color=black, line width=0.6pt},
			y tick style={color=black, line width=0.6pt},
			ytick pos=left,
			major tick length=1mm,
			minor y tick num=4,
			minor y tick style={color=gray, line width=0.4pt},
			minor tick length=0.75mm,
			grid style={dashed, lightgray, dash pattern=on 1pt off 1pt},
			]
			
			\addplot[name path=Smean, thick, set2cyan]
			table [x=TimeFound, y expr=\thisrow{mean}/10.269, col sep=comma] {figures/SANJESH.csv};
			\addlegendentry{\sysname}
			
			\addplot[name path=Supper, draw=none, forget plot]
			table [x=TimeFound, y expr=(\thisrow{mean} + \thisrow{std})/10.269, col sep=comma] {figures/SANJESH.csv};
			\addplot[name path=Slower, draw=none, forget plot]
			table [x=TimeFound, y expr=(\thisrow{mean} - \thisrow{std})/10.269, col sep=comma] {figures/SANJESH.csv};
			\addplot[set2cyan, opacity=0.15, forget plot] fill between[of=Supper and Slower];
			
			\addplot[name path=GAmean, thick, set2green]
			table [x=TimeFound, y expr=\thisrow{mean}/16.33, col sep=comma] {figures/Gap-Progress/Genetics_400_Gap_Progress.csv};
			\addlegendentry{\genetics}
			
			\addplot[name path=GAupper, draw=none, forget plot]
			table [x=TimeFound, y expr=(\thisrow{mean} + \thisrow{std})/16.33, col sep=comma] {figures/Gap-Progress/Genetics_400_Gap_Progress.csv};
			\addplot[name path=GAlower, draw=none, forget plot]
			table [x=TimeFound, y expr=(\thisrow{mean} - \thisrow{std})/16.33, col sep=comma] {figures/Gap-Progress/Genetics_400_Gap_Progress.csv};
			\addplot[set2green, opacity=0.12, forget plot] fill between[of=GAupper and GAlower];
			
			\addplot[name path=SAmean, thick, set2blue]
			table [x=TimeFound, y expr=\thisrow{mean}/16.33, col sep=comma] {figures/Gap-Progress/SA_400_Gap_Progress.csv};
			\addlegendentry{\sa}
			
			\addplot[name path=SAupper, draw=none, forget plot]
			table [x=TimeFound, y expr=(\thisrow{mean} + \thisrow{std})/16.33, col sep=comma] {figures/Gap-Progress/SA_400_Gap_Progress.csv};
			\addplot[name path=SAlower, draw=none, forget plot]
			table [x=TimeFound, y expr=(\thisrow{mean} - \thisrow{std})/16.33, col sep=comma] {figures/Gap-Progress/SA_400_Gap_Progress.csv};
			\addplot[set2blue, opacity=0.12, forget plot] fill between[of=SAupper and SAlower];
			
			\addplot[name path=HCmean, thick, set2purple]
			table [x=TimeFound, y expr=\thisrow{mean}/16.33, col sep=comma] {figures/Gap-Progress/HillClimbing_400_Gap_Progress.csv};
			\addlegendentry{\hill}
			
			\addplot[name path=HCupper, draw=none, forget plot]
			table [x=TimeFound, y expr=(\thisrow{mean} + \thisrow{std})/16.33, col sep=comma] {figures/Gap-Progress/HillClimbing_400_Gap_Progress.csv};
			\addplot[name path=HClower, draw=none, forget plot]
			table [x=TimeFound, y expr=(\thisrow{mean} - \thisrow{std})/16.33, col sep=comma] {figures/Gap-Progress/HillClimbing_400_Gap_Progress.csv};
			\addplot[set2purple, opacity=0.12, forget plot] fill between[of=HCupper and HClower];
			
			\addplot[name path=Simmean, thick, set2brown]
			table [x=TimeFound, y expr=\thisrow{mean}/16.33, col sep=comma] {figures/Gap-Progress/Simulation_400_Gap_Progress.csv};
			\addlegendentry{\simul}
			
			\addplot[name path=Simupper, draw=none, forget plot]
			table [x=TimeFound, y expr=(\thisrow{mean} + \thisrow{std})/16.33, col sep=comma] {figures/Gap-Progress/Simulation_400_Gap_Progress.csv};
			\addplot[name path=Simlower, draw=none, forget plot]
			table [x=TimeFound, y expr=(\thisrow{mean} - \thisrow{std})/16.33, col sep=comma] {figures/Gap-Progress/Simulation_400_Gap_Progress.csv};
			\addplot[set2brown, opacity=0.12, forget plot] fill between[of=Simupper and Simlower];
			
			\nextgroupplot[
			width=0.42\textwidth,
			boxplot/draw direction=y,
			xtick={1,2,3,4,5},
			xticklabels={
				\small\sysname,
				\small \genetics,
				\small \sa,
				\small \hill,
				\small \simul,
			},
			xtick align=outside,
			xtick pos=bottom,
			x tick style={color=black, line width=0.6pt},
			y tick style={color=black, line width=0.6pt},
			ytick pos=left,
			major tick length=1mm,
			ymajorgrids=true,
			yminorgrids=false,
			xmajorgrids=false,
			grid style={dashed, lightgray, dash pattern=on 1pt off 1pt},
			]
			
			\addplot+[
			boxplot prepared={
				median=1.0000,
				lower quartile=0.93,
				upper quartile=1.06,
				upper whisker=1.22,
				lower whisker=0.75
			},
			fill=set2cyan,
			draw=set2gray,
			line width=0.75pt
			] coordinates {};
			
			\addplot+[
			boxplot prepared={
				median=0.61,
				lower quartile=0.55,
				upper quartile=0.66,
				upper whisker=0.79,
				lower whisker=0.39
			},
			fill=set2green,
			draw=set2gray,
			line width=0.75pt
			] coordinates {};
			
			\addplot+[
			only marks, draw=set2gray, fill=none, mark=o, mark size=1.5, line width=0.7pt
			] coordinates {(2, 0.38)};
			
			\addplot+[
			boxplot prepared={
				median=0.38,
				lower quartile=0.31,
				upper quartile=0.47,
				upper whisker=0.67,
				lower whisker=0.19
			},
			fill=set2blue,
			draw=set2gray,
			line width=0.75pt
			] coordinates {};
			
			\addplot+[
			boxplot prepared={
				median=0.24,
				lower quartile=0.23,
				upper quartile=0.26,
				upper whisker=0.32,
				lower whisker=0.17
			},
			fill=set2purple,
			draw=set2gray,
			line width=0.75pt
			] coordinates {};
			
			\addplot+[
			only marks, draw=set2gray, fill=none, mark=o, mark size=1.5, line width=0.7pt
			] coordinates {(4, 0.091)};
			\addplot+[
			only marks, draw=set2gray, fill=none, mark=o, mark size=1.5, line width=0.7pt
			] coordinates {(4, 0.13)};
			\addplot+[
			only marks, draw=set2gray, fill=none, mark=o, mark size=1.5, line width=0.7pt
			] coordinates {(4, 0.15)};
			\addplot+[
			only marks, draw=set2gray, fill=none, mark=o, mark size=1.5, line width=0.7pt
			] coordinates {(4, 0.10)};
			\addplot+[
			only marks, draw=set2gray, fill=none, mark=o, mark size=1.5, line width=0.7pt
			] coordinates {(4, 0.325)};
			\addplot+[
			only marks, draw=set2gray, fill=none, mark=o, mark size=1.5, line width=0.7pt
			] coordinates {(4, 0.327)};
			
			\addplot+[
			boxplot prepared={
				median=0.25,
				upper quartile=0.27,
				lower quartile=0.23,
				upper whisker=0.27,
				lower whisker=0.22
			},
			fill=set2brown,
			draw=set2gray,
			line width=0.75pt
			] coordinates {};
		\end{groupplot}
	\end{tikzpicture}
	
	\vspace{-1em}
	\caption{\sysname vs. baselines. Left: risk of migration over time (normalized to \sysname mean). \sysname increases risk as it analyzes partitions and uncovers high-risk scenarios early, while baselines improve slowly. Right: distribution of final risk across runs. \sysname finds up to $\sim 4\times$ higher risk than \simul and exceeds the best baseline (\genetics) by $1.6\times$ in median.}
	\label{fig:sanjeshvsimul}
\end{figure*}

\subsection{\sysname Outperforms Benchmarks}
\label{sec:query_1}

\cref{fig:sanjeshvsimul} (right) shows \sysname and the baselines on query~\ref{query:1}. \sysname finds scenarios with up to $\sim 4\times$ higher migration risk than the operator’s evaluator (\simul). Among search-based baselines, \sysname exceeds \genetics by $1.6\times$ in median. Importantly, $46\%$ of the high-risk feature combinations \sysname uncovers appear in the production traces---these are not artificial scenarios.

The baselines sample from production traces and explore only a small portion of the combinatorial search space. They miss interactions between VM sequences, ML predictions, and system dynamics that compound into high-risk scenarios.

\cref{fig:sanjeshvsimul} (left) shows how risk evolves over time as each method searches for adversarial sequences. \sysname increases risk as it analyzes successive partitions and uncovers high-risk patterns early. The baselines operate on full sequences and improve slowly, which prevents them from reaching similar risk levels. \sysname reaches the same risk as \genetics using $\sim\!1/3$ of the sequence on average, which shows that it identifies critical patterns more efficiently.

We run this experiment with sequences of 400 VMs to highlight the performance gap across methods, while the remaining experiments use $N=200$. 

We also evaluate the necessity of time partitioning by running Phase~1 of \sysname without partitioning on a sequence of 200 VMs. After 24 hours, the solver does not produce any adversarial predictions, which shows that directly solving the bi-level optimization at this scale is intractable.




\graybox{\textbf{How operators can use \sysname's output.}
	
	1. Add \sysname-generated scenarios to the allocator’s automated regression suite so updated ML model versions are continuously validated against these high-risk placement cases.\\
	2. Retrain ML models with increased weight on training samples that match \sysname’s scenarios, which improves robustness in the vulnerable regions of the workload/feature space.}


{\footnotesize
\begin{figure}[t]
	\centering
	\begin{tikzpicture}
		\begin{axis}[
			boxplotstyle,
			width=0.95\columnwidth,
			height=0.5\columnwidth,
			xtick={1,2,3,4},
			xticklabels={
				\textsc{S}, 
				\textsc{S}\textsubscript{\textsc{M}}, 
				\textsc{S}\textsubscript{\textsc{C}}, 
				\textsc{S}\textsubscript{\textsc{L}}},
			enlarge x limits=0.05,
			ymin=-0.1, ymax=1.5,
			ylabel={\small Risk of Migration},
			ylabel style={font=\large},       
			yticklabel style={font=\large},   
			boxplot/box extend=0.7,
			]
			
            \addplot+[
        	boxplot prepared={
        		median=1.0000,
        		lower quartile=0.9081,
        		upper quartile=1.0753,
        		upper whisker=1.3249,
        		lower whisker=0.6883
        	},
        	fill=set2cyan,
        	draw=set2gray,
        	line width=0.75pt
        	] coordinates {};

            \addplot+[
        	only marks, draw=set2gray, fill=none, mark=o, mark size=1.5, line width=0.7pt
        	] coordinates {(1, 0.6289)};
			
			\addplot+[
			boxplot prepared={
				median=0.1260,
				upper quartile=0.1605,
				lower quartile=0.0880,
				upper whisker=0.2588,
				lower whisker=0.0154
			},
			fill=set2orange,
			draw=set2gray,
			line width=0.75pt
			] coordinates {};
			
			\addplot+[
			boxplot prepared={
				median=0.9430,
				upper quartile=1.0000,
				lower quartile=0.8773,
				upper whisker=1.1816,
				lower whisker=0.7154
			},
			fill=set2blue,
			draw=set2gray,
			line width=0.75pt
			] coordinates {};
			
			\addplot+[
			only marks, draw=set2gray, fill=none, mark=o, mark size=1.5, line width=0.7pt
			] coordinates {
				(3, 1.1942)
			};
			
			\addplot+[
			boxplot prepared={
				median=0,
				upper quartile=0,
				lower quartile=0,
				upper whisker=0,
				lower whisker=0
			},
			fill=set2purple,
			draw=set2gray,
			line width=0.75pt
			] coordinates {};
		\end{axis}
	\end{tikzpicture}
	\caption{Impact of each model on the risk of migration. The CPU model contributes the most, driving up to $6\times$ higher migration risk compared to the memory model. The lifetime model has no impact when the others are accurate. The x-axis label \textsc{S} denotes \sysname; subscripts indicate which model provides predictions while the others use ground truth (\textsc{M}: memory, \textsc{C}: CPU, \textsc{L}: lifetime)}
	\label{fig:system-interaction}
\end{figure}
}

\subsection{The Models' Interaction With the System}
\label{sec:query_2}

\autoref{fig:system-interaction} shows how the models affect the VM allocator differently (query~\ref{query:2}). In this experiment, the allocator uses the memory model together with the CPU and lifetime models to make placement decisions. When all three models provide predictions, \sysname finds scenarios with a $6\%$ higher risk of migration (median) compared to using only the CPU model’s predictions with ground-truth labels for the others ($\text{\sysname}_{C}$). This $6\%$ gap is the measurable effect of model interactions: it represents additional over-utilized servers that arise solely from how memory and lifetime predictions compound with CPU errors through the bin-packing heuristic---invisible to any per-model analysis. 

The CPU model has the largest impact: $\text{\sysname}_{C}$ causes a $6\times$ higher risk of migration compared to $\text{\sysname}_{M}$ (where the system uses only predictions from the memory model).
The reasons for this higher impact are (1) a single VM can consume an entire server’s CPU but not its memory; and (2) the CPU model is less granular than the memory model which means its mispredictions are more likely to cause the allocator to place the VM on a server with insufficient capacity. 
The lifetime model alone ($\text{\sysname}_{L}$) has no impact on risk of migration when the other models are accurate---a finding that tells operators they can deprioritize engineering effort on the lifetime model for migration-risk reduction. 

\graybox{\textbf{How operators can use \sysname's output.}

\sysname demonstrates that the CPU model dominates the impact on end-to-end performance. Operators can act on this insight in two ways: (i) prioritize improving the CPU model’s accuracy, or (ii) apply an asymmetric safety margin by inflating CPU predictions with an $x\%$ buffer (i.e., treat each VM as needing $x\%$ more CPU than predicted) before bin-packing. Predictions from the other models can be used as-is.}


\subsection{\sysname Analyzes Data Drift}
\label{sec:query_3}
\sysname analyzes how data drift affects system performance (query~\ref{query:3}, \autoref{fig:drift}).
We simulate $20\%$ and $40\%$ drift by changing the feature-to-label mapping for that fraction of the data. \autoref{fig:drift} shows that drift in the lifetime model increases migration risk more than drift in the CPU model: at $40\%$ drift, risk rises by $7\%$ for the lifetime model but only $3\%$ for the CPU model. The difference between $20\%$ and $40\%$ drift is not statistically significant for either model, indicating a saturation effect. Overall, the VM allocator is robust to moderate data drift.

\graybox{\textbf{How operators can use \sysname's output.}

\sysname shows that operators can safely reduce the CPU model’s retraining cadence (e.g., from daily to weekly), which cuts training-related compute costs without increasing operational risk.}

\begin{figure}[t]
\centering
\begin{tikzpicture}
\begin{axis}[
	boxplotstyle,
	boxplot/box extend=0.20,   
	width=1\columnwidth,
	height=0.53\columnwidth,
	xtick={1,2,3},
	xticklabels={
		\small\textsc{\sysname}, 
		\small\textsc{\sysname}\textsubscript{\scriptsize$20\%$Drift}, 
		\small\textsc{Sanjesh}\textsubscript{\scriptsize$40\%$Drift}
	},
	enlarge x limits=0.05,
	ylabel={\small Risk of Migration},
	ylabel style={font=\large},
	yticklabel style={font=\large},
	ymin=0.5, ymax=1.5,
	clip=false
	]
	\def\dx{0.15} 
	
	\addplot+[
	boxplot prepared={
		median=1.0000,
		lower quartile=0.9081,
		upper quartile=1.0753,
		upper whisker=1.3249,
		lower whisker=0.6883
	},
	fill=set2cyan,
	draw=set2gray,
	line width=0.75pt
	] coordinates {};
	
	\addplot+[ 
	boxplot prepared={
		median=1.0045,
		upper quartile=1.0900,
		lower quartile=0.9142,
		upper whisker=1.3364,
		lower whisker=0.7076
	},
	boxplot/draw position={2-\dx},
	fill=set2purple,
	draw=set2gray,
	line width=0.75pt
	] coordinates {};
	\addplot+[ 
	boxplot prepared={
		median=1.0832,
		upper quartile=1.1551,
		lower quartile=0.9890,
		upper whisker=1.3891,
		lower whisker=0.8240
	},
	boxplot/draw position={2+\dx},
	fill=set2blue,
	draw=set2gray,
	line width=0.75pt
	] coordinates {};
	
	\addplot+[ 
	boxplot prepared={
		median=1.0318,
		upper quartile=1.1104,
		lower quartile=0.9339,
		upper whisker=1.3566,
		lower whisker=0.7542
	},
	boxplot/draw position={3-\dx},
	fill=set2purple,
	draw=set2gray,
	line width=0.75pt
	] coordinates {};
	\addplot+[ 
	boxplot prepared={
		median=1.0790,
		upper quartile=1.1596,
		lower quartile=0.9940,
		upper whisker=1.3852,
		lower whisker=0.8102
	},
	boxplot/draw position={3+\dx},
	fill=set2blue,
	draw=set2gray,
	line width=0.75pt
	] coordinates {};
	
	\addplot+[ 
	only marks, draw=set2gray, fill=none, mark=o, mark size=1.5, line width=0.7pt
	] coordinates {(1, 0.6289)};
	\addplot+[ 
	only marks, draw=set2gray, fill=none, mark=o, mark size=1.5, line width=0.7pt
	] coordinates {(2-\dx, 1.3608)};
	\addplot+[ 
	only marks, draw=set2gray, fill=none, mark=o, mark size=1.5, line width=0.7pt
	] coordinates {(2-\dx, 1.3929)};
	\addplot+[ 
	only marks, draw=set2gray, fill=none, mark=o, mark size=1.5, line width=0.7pt
	] coordinates {(2+\dx, 1.48)};
	\addplot+[ 
	only marks, draw=set2gray, fill=none, mark=o, mark size=1.5, line width=0.7pt
	] coordinates {(3+\dx, 1.44)};
	\node at (axis cs:2-\dx,0.78) [anchor=north,yshift=-5pt] {\scriptsize$D_C$};
	\node at (axis cs:2+\dx,0.78) [anchor=north,yshift=-5pt] {\scriptsize$D_L$};
	
	\node at (axis cs:3-\dx,0.78) [anchor=north,yshift=-5pt] {\scriptsize$D_C$};
	\node at (axis cs:3+\dx,0.78) [anchor=north,yshift=-5pt] {\scriptsize$D_L$};
\end{axis}
\end{tikzpicture}
\caption{$D_C$ and $D_L$ denote drift in CPU and Lifetime model data. Introducing $20\%$ and $40\%$ drift causes only a small increase in migration risk. This shows that the VM allocator is robust to moderate drift.}
\label{fig:drift}
\end{figure}

\subsection{\sysname Analyzes a Hypothetical Model}
\label{sec:query_9}
Given the CPU model’s outsized impact (\autoref{fig:system-interaction}), we investigate how much end-to-end performance improves with a more accurate CPU model (query~\ref{query:5}). \autoref{fig:hypothetical} shows that a $10\%$ accuracy improvement---while keeping $90\%$ of predictions unchanged---cuts migration risk by $2\times$. Operators can further constrain the hypothetical model’s error profile (e.g., bias improvements toward reducing underprediction) to identify which changes deliver the greatest payoff and what additional data to prioritize collecting (similar to~\cite{pants}).


\graybox{\textbf{How operators can use \sysname's output.}

The non-linear payoff, where a 10\% accuracy improvement translates to a 50\% reduction in risk, provides a financial justification for investing engineering effort into improving the CPU model.}
\begin{figure}[t]
\centering
\begin{tikzpicture}
\begin{axis}[
	boxplotstyle,
    width=0.7\columnwidth,
	height=0.45\columnwidth,
	xticklabels={\sysname,
		\sysname\textsubscript{{CPUBoost}}},
	ylabel={\small Risk of Migration},
	ylabel style={font=\large},
	yticklabel style={font=\large},
	ymin=0, ymax=1.5
	]
	
	\addplot+[
	boxplot prepared={
		median=1.0000,
		lower quartile=0.9081,
		upper quartile=1.0753,
		upper whisker=1.3249,
		lower whisker=0.6883
	},
	fill=set2cyan,
	draw=set2gray,
	line width=0.75pt
	] coordinates {};
	
	\addplot+[
	boxplot prepared={
		median=0.4047,
		upper quartile=0.5549,
		lower quartile=0.3816,
		upper whisker=0.6121,
		lower whisker=0.3802
	},
	fill=set2orange,
	draw=set2gray,
	line width=0.75pt
	] coordinates {};
	
	\addplot+[
	only marks, draw=set2gray, fill=none, mark=o, 
	mark size=1.5, line width=0.7pt
	] coordinates {(1, 0.6289)};
	
\end{axis}
\end{tikzpicture}
\caption{\sysname quantifies the improvement we get if we improve the CPU model by $10\%$.}
\label{fig:hypothetical}
\end{figure}

\begin{figure}[t]
\centering
\begin{tikzpicture}
    \begin{axis}[
            boxplotstyle,
            width=0.65 \columnwidth,
            height=0.55 \columnwidth,
            xticklabels={\sysname, \simul},
            ylabel={\Large $\frac{\text{Num~Excess~Servers}}{\text{Epoch}}$},
            ytick={0,5,10,15,20,25},
            ymax=25
        ]
        
            \addplot+[
                boxplot prepared={
                        median=16.8897,
                        lower quartile=13.6328,
                        upper quartile=18.9983,
                        upper whisker=24,
                        lower whisker=6.2828
                    },
                fill=set2cyan,
                draw=set2gray,
                line width=0.75pt
            ] coordinates {};
        
            \addplot+[
                boxplot prepared={
                        median=2.1655,
                        upper quartile=2.2138,
                        lower quartile=1.8759,
                        upper whisker=2.2414,
                        lower whisker=1.8276
                    },
                fill=set2orange,
                draw=set2gray,
                line width=0.75pt
            ] coordinates {};
        
        \end{axis}
\end{tikzpicture}
\caption{\sysname vs. \simul on server utilization. \sysname finds scenarios with up to $8\times$ more excess servers per epoch (relative to the oracle) than \simul. This gap shows that \simul significantly underestimates worst-case resource overhead.}
\label{fig:server-util}
\end{figure}

\begin{figure*}[t]
\centering

\begin{minipage}[t]{0.32\textwidth}
\centering
\resizebox{\linewidth}{!}{%
\begin{tikzpicture}
	\def\barheight{0.4}
	\def\spacing{0.5}
	\def\xscale{4.25}
	
	\node[anchor=east] at (0, 0*\spacing) {\small Feature\textsubscript{A}};
	\node[anchor=east] at (0, 1*\spacing) {\small Feature\textsubscript{B}};
	\node[anchor=east] at (0, 2*\spacing) {\small Feature\textsubscript{C}};
	\node[anchor=east] at (0, 3*\spacing) {\small Feature\textsubscript{E}};
	
	\foreach \xstart/\xend in {
		0.0/0.02294,
		0.02294/0.03837,
		0.03837/0.05662,
		0.05662/0.06227,
		0.06227/0.08672,
		0.37594/0.39743,
		0.48098/0.48568,
		0.66595/0.70181,
		0.70181/0.72140,
		0.76258/0.88315,
		0.88315/1.0
	} {
		\draw[fill=shapblue, draw=darkgray]
		(\xscale*\xstart, 0*\spacing - \barheight/2) rectangle
		(\xscale*\xend,   0*\spacing + \barheight/2);
	}
	
	\foreach \y in {1,2,3} {
		\draw[fill=shapred!60, draw=darkgray]
		(0, \y*\spacing - \barheight/2) rectangle
		(\xscale*1.0, \y*\spacing + \barheight/2);
	}
	
	\draw[->] (0, -\barheight) -- (\xscale*1.05, -\barheight)
	node[below, yshift=-0.45cm, xshift=-2.5cm] {\small Normalized Feature Range};
	\foreach \x in {0, 0.2, 0.4, 0.6, 0.8, 1.0} {
		\draw (\xscale*\x, -\barheight) -- (\xscale*\x, -\barheight - 0.1)
		node[below] {\small \x};
	}
\end{tikzpicture}%
}
\end{minipage}\hfill%
\begin{minipage}[t]{0.32\textwidth}
\centering
\resizebox{\linewidth}{!}{%
\begin{tikzpicture}
	\def\barheight{0.4}
	\def\spacing{0.5}
	\def\xscale{4.25}
	
	\node[anchor=east] at (0, 0*\spacing) {\small Feature\textsubscript{A}};
	\node[anchor=east] at (0, 1*\spacing) {\small Feature\textsubscript{B}};
	\node[anchor=east] at (0, 2*\spacing) {\small Feature\textsubscript{C}};
	\node[anchor=east] at (0, 3*\spacing) {\small Feature\textsubscript{E}};
	
	\foreach \xstart/\xend in {0.0/0.022942494180755505} {
		\draw[fill=shapblue, draw=darkgray]
		(\xscale*\xstart, 0*\spacing - \barheight/2) rectangle
		(\xscale*\xend,   0*\spacing + \barheight/2);
	}
	
	\foreach \y in {1,2,3} {
		\draw[fill=shapred!60, draw=darkgray]
		(0, \y*\spacing - \barheight/2) rectangle
		(\xscale*1.0, \y*\spacing + \barheight/2);
	}
	
	\draw[->] (0, -\barheight) -- (\xscale*1.05, -\barheight)
	node[below, yshift=-0.45cm, xshift=-2.5cm] {\small Normalized Feature Range};
	\foreach \x in {0, 0.2, 0.4, 0.6, 0.8, 1.0} {
		\draw (\xscale*\x, -\barheight) -- (\xscale*\x, -\barheight - 0.1)
		node[below] {\small \x};
	}
\end{tikzpicture}%
}
\end{minipage}\hfill%
\begin{minipage}[t]{0.32\textwidth}
\centering
\includegraphics[width=\linewidth]{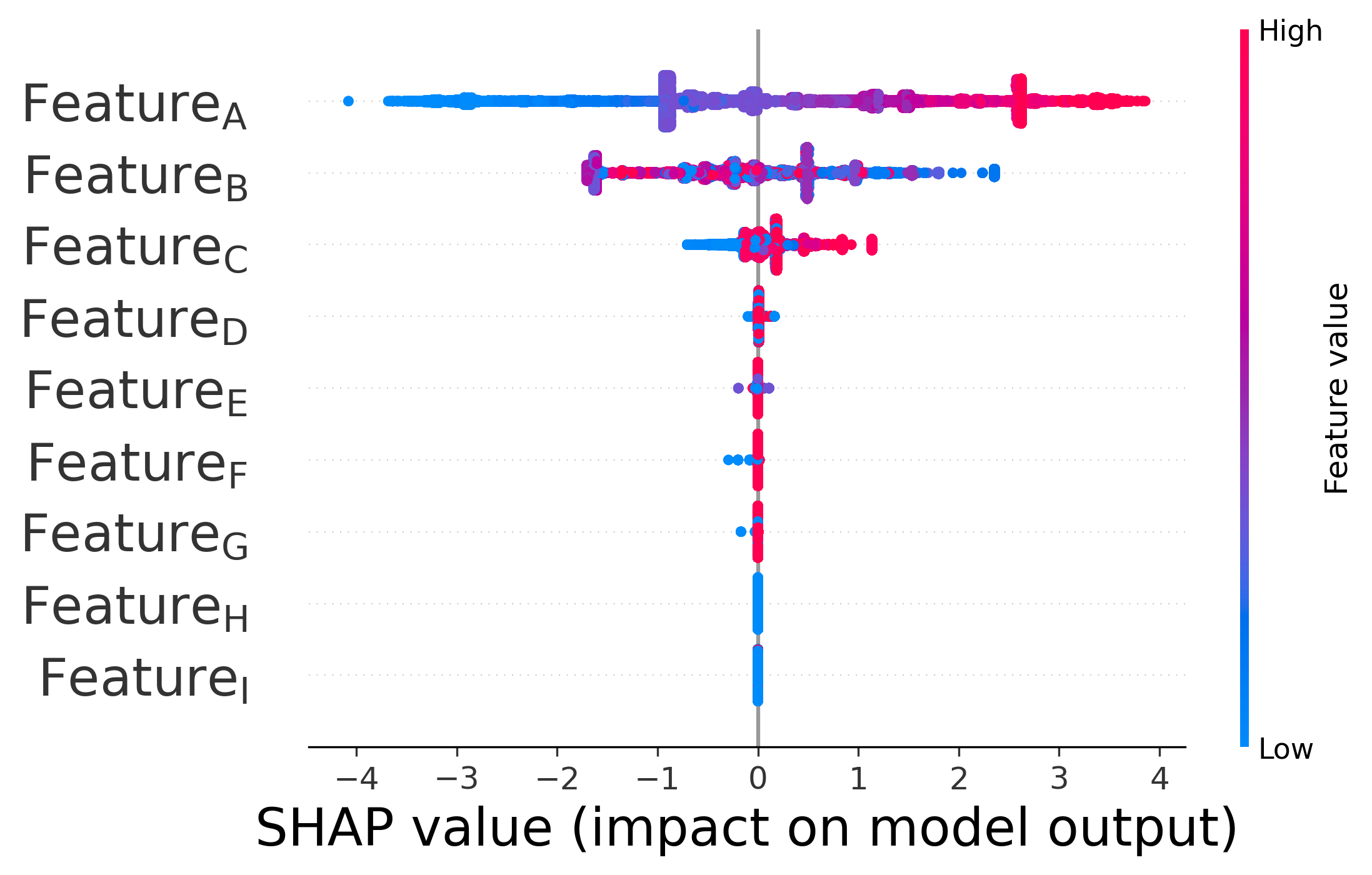}
\end{minipage}

\caption{\sysname produces a risk surface that aligns with SHAP. (Left) shows feature risk regions across 50+ VMs, (middle) across 100+ VMs, and (right) shows SHAP values. While SHAP ranks features by their influence on model predictions, \sysname highlights regions (in blue) that cause poor system performance and marks high-entropy, non-actionable features in red. Both methods identify feature A as the most important that strongly affects both model predictions and system performance.}
\label{fig:risk_regions}
\end{figure*}

\subsection{\sysname Analyzes Server Utilization}
\sysname supports multiple objectives as long as they can be expressed as linear constraints in \autoref{eq:bi-level} (see \cref{sec:objective-formulation} for formulations of different objectives). In this experiment, we change the objective to maximize the gap in the number of servers used by the allocator with ML predictions and the oracle with access to ground-truth labels, and compare the resulting scenarios against the operator’s evaluator (\simul).

\autoref{fig:server-util} shows that the allocator can use up to 16 more servers per epoch on average than the oracle. \simul fails to uncover such scenarios and instead finds cases with only 2 extra servers per epoch on average.

\graybox{\textbf{How operators can use \sysname's output.}

Reserve spare capacity based on worst-case headroom rather than the simulation average (up to $8\times$ more servers) to avoid under-provisioning.}

\begin{figure*}[t]
\centering
\includegraphics[width=0.7\textwidth]{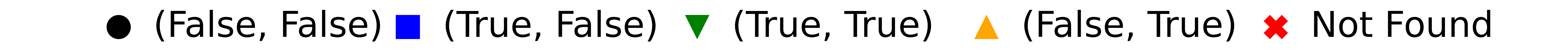}

\vspace{1.5ex}

\begin{minipage}[t]{0.33\textwidth}
  \centering
  \includegraphics[width=\linewidth]{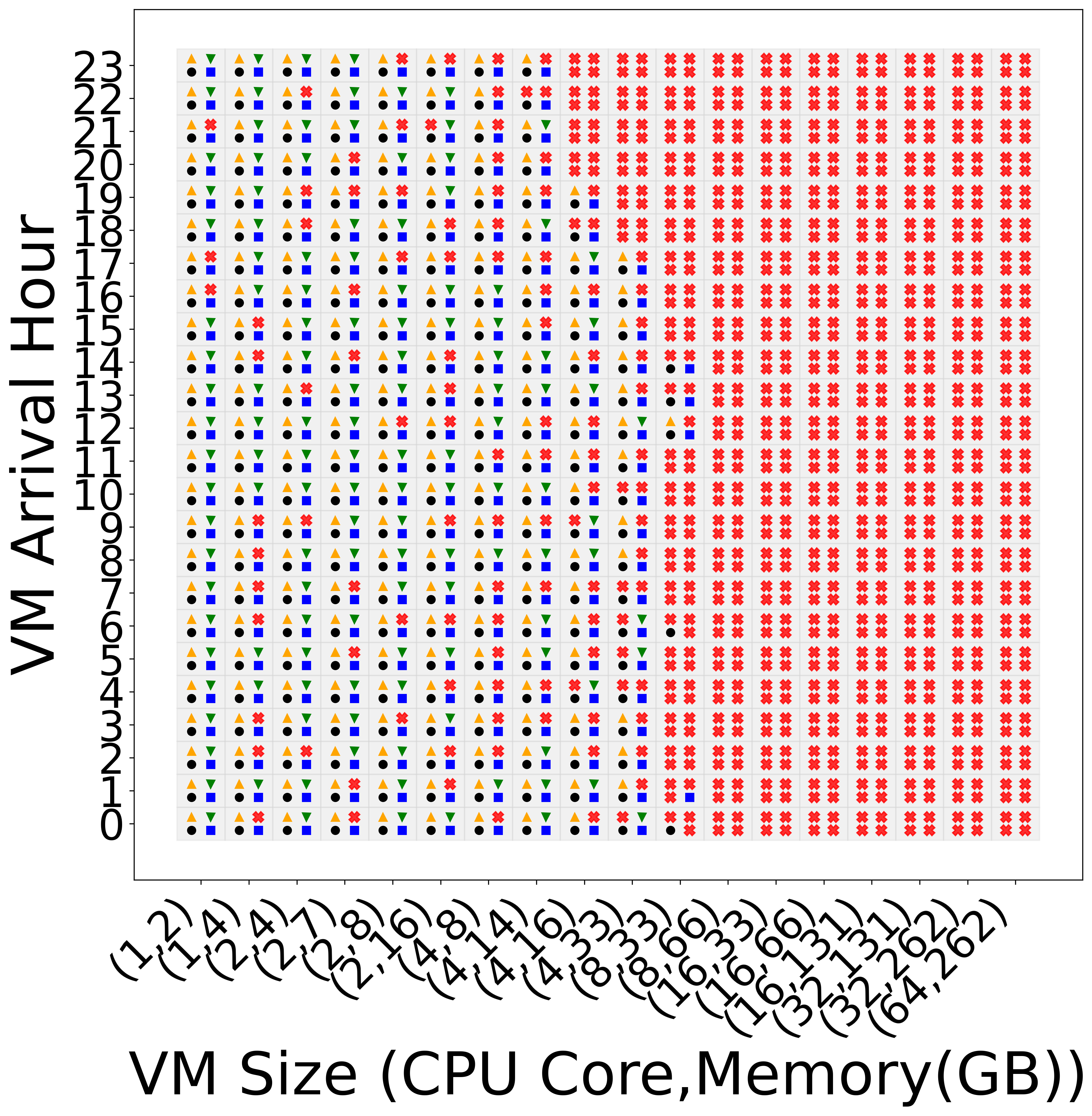}
  \vspace{2pt}
  
  {\small (a) CEGAR after 1 min}
\end{minipage}\hfill
\begin{minipage}[t]{0.33\textwidth}
  \centering
  \includegraphics[width=\linewidth]{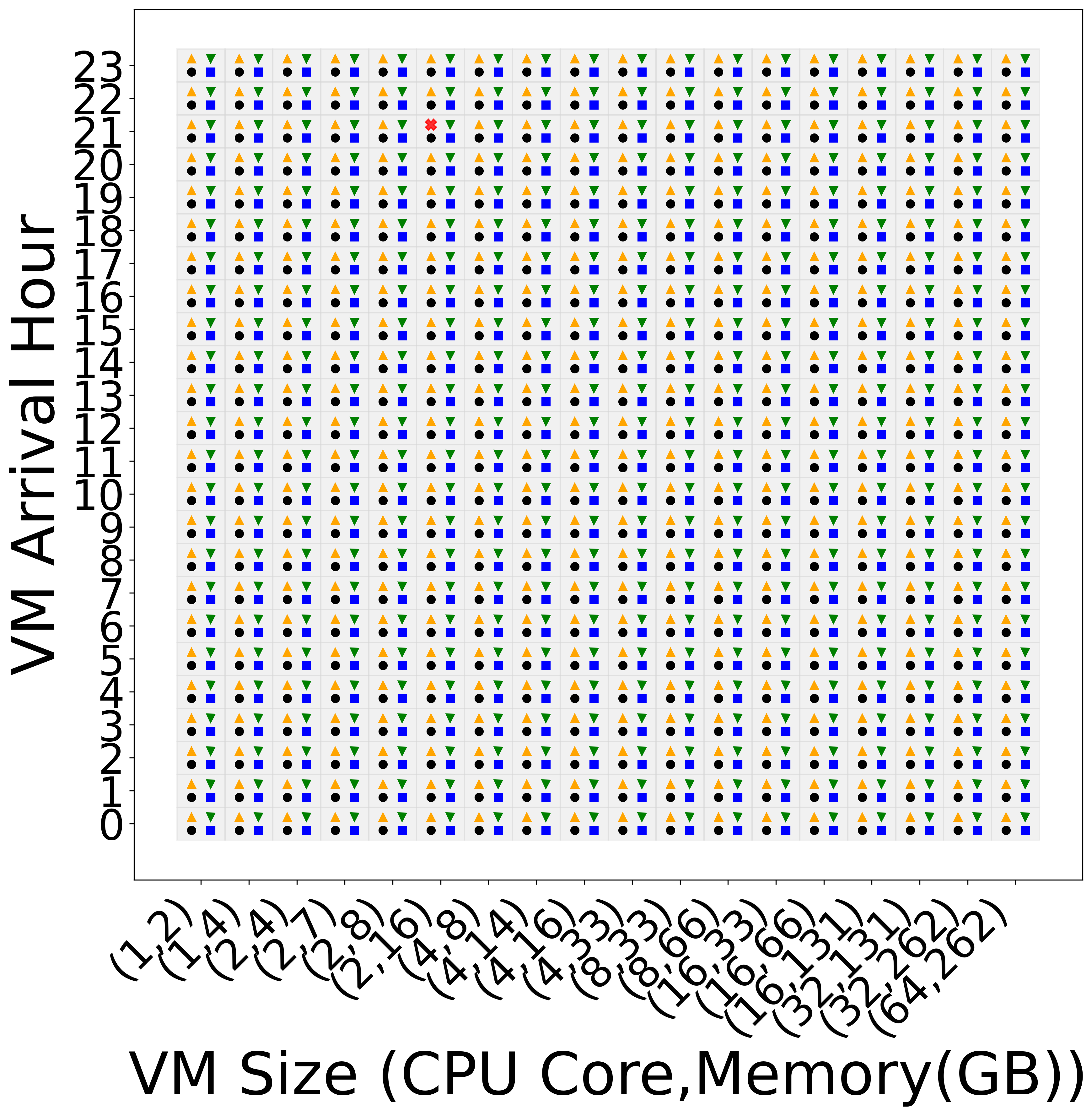}
  \vspace{2pt}
  
  {\small (b) CEGAR after 5 min}
\end{minipage}\hfill
\begin{minipage}[t]{0.33\textwidth}
  \centering
  \includegraphics[width=\linewidth]{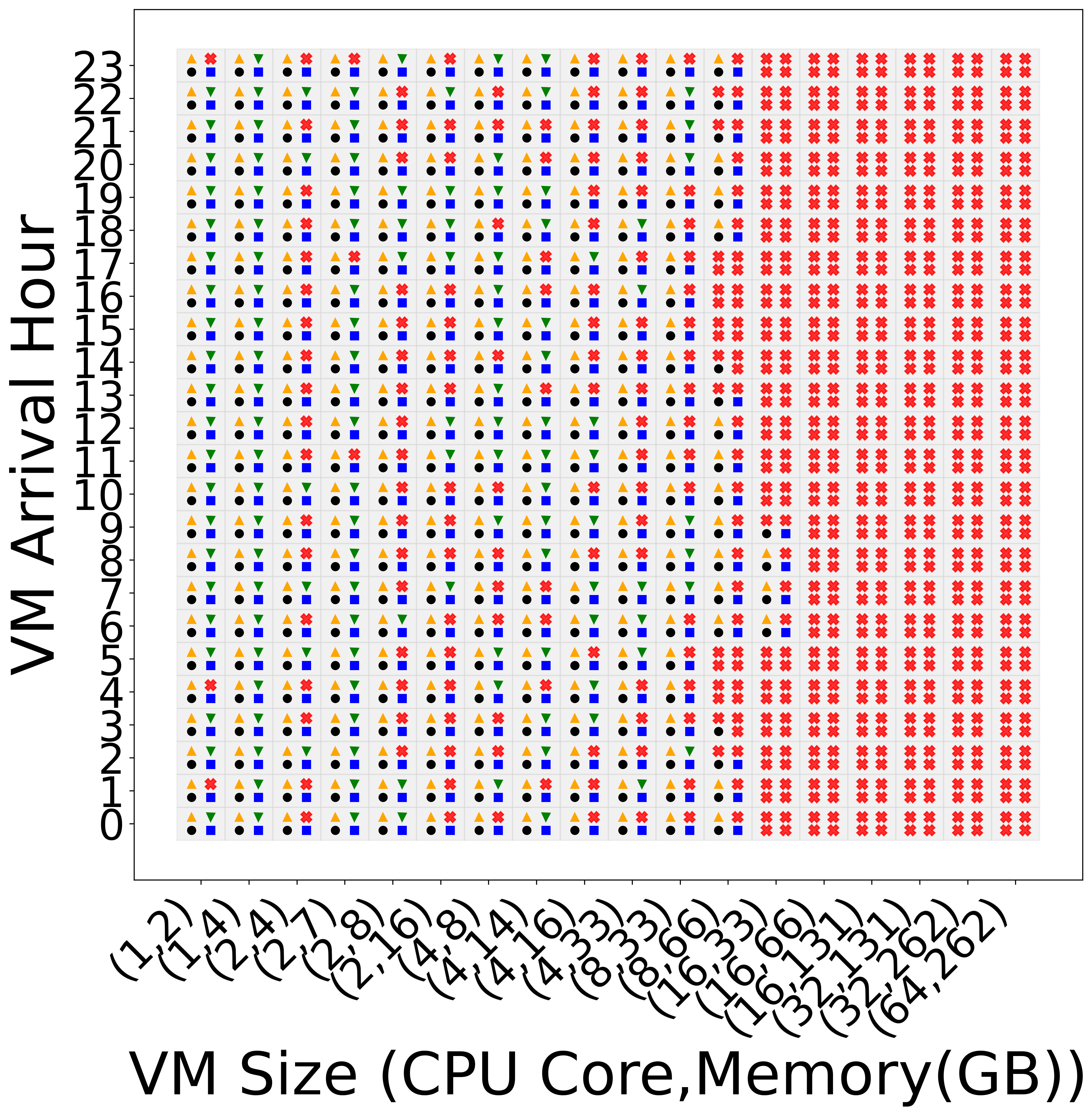}
  \vspace{2pt}
  
  {\small (c) No CEGAR after 5 min}
\end{minipage}
\caption{Phase~2 runtime analysis. Each cell indicates whether a feasible feature vector has been found for a given pair of model output labels at a specific VM resource size and arrival hour. CEGAR-based approach finds all combinations within one hour (a,\,b); without CEGAR-based approach, direct Z3 encoding finds only 57\% in 5 minutes~(c).}
\label{fig:cegar_performance_analysis}
\end{figure*}

\subsection{\sysname Computes Risk Surface}
\label{sec:shap}
Given the CPU model’s dominance, we use \sysname to identify \emph{where} in feature space it causes performance degradation (query~\ref{query:4}). SHAP~\cite{shap} highlights which features influence the CPU model’s predictions---e.g., identifying ``feature A'' as highly important---but does not connect those features to system-level outcomes. \sysname’s post-analysis (\cref{sec:risk_surface_main}) goes further: it shows that ``feature A'' leads to poor \emph{system} performance only within specific feature ranges (\autoref{fig:risk_regions}), and lets operators define their risk tolerance to control how aggressively the guardrail triggers.

\graybox{\textbf{How operators can use \sysname's output.}

Implement a guardrail policy for the CPU model: 
\textsc{IF $Feature_A \in [0, 0.1]$ THEN bypass the model and use a conservative allocation}. 
This mitigates risk without retraining and enables an immediate production patch for the exposed vulnerability.}

\subsection{Phase~2 Scalability}
\label{sec:phase2_perf}


\cref{fig:cegar_performance_analysis} shows the performance of our CEGAR-based approach in Phase~2.
Our approach tries to find the features for all combinations of binary outputs of the CPU and lifetime models across different VM sizes and VM arrival hours within a day. VM size and arrival hour are the two features that \sysname determines in Phase~1 alongside the model predictions.

The abstraction--refinement loop identifies all feasible feature combinations within one hour and recovers all but one within five minutes. In contrast, directly encoding the full tree structure in Z3 recovers only $57\%$ within the same time budget.
Early detection of infeasible combinations is critical. When no feature vector exists for a prediction pattern from Phase~1, we must exclude that pattern and re-run Phase~1. Our approach enables this pruning quickly, avoiding wasted computation.

\section{Discussion}
\label{sec:discussion}

\sysname is the first system that analyzes how predictions from \emph{multiple} ML models jointly affect end-to-end system performance.
Through a case study of a large public cloud’s VM allocator~\cite{RC}, we demonstrated the range of queries \sysname can answer. Beyond the scenarios we evaluated:

\parab{\sysname is general.} Phase~1 applies to any ML-augmented system whose logic can be modeled as a Stackelberg game (\autoref{eq:bi-level})---concretely, any system whose algorithms can be expressed as a feasibility or convex optimization problem. Our implementation already supports ML-based caching~\cite{danielcaching}\footnote{We did not find public data for ML-based caching~\cite{danielcaching}, but our implementation natively supports it.} and traffic engineering~\cite{dote}; details in \cref{sec:extensibility}.


MetaOpt~\cite{metaopt} introduces topology-based partitioning. Combined with our time-partitioning technique (\cref{sec:time-partitioning}), \sysname scales Phase~1 to systems with thousands of decision variables. For systems with LGBMs, Phase~2 uses a CEGAR-based approach that scales to deep and dense models through abstraction--refinement, avoiding the limitations of direct SMT encodings. \cref{sec:time-partitioning} shows that our approach finds $99.9\%$ of feasible feature combinations within 5 minutes, compared to $57\%$ coverage without abstraction under the same time budget.

\parab{\sysname applies to non-LGBM models.} Our two-phase decomposition---first identifying harmful predictions, then mapping them to features---means Phase~2 can analyze any model for which we can express the prediction-to-feature mapping as constraints. \cite{namyar2024end} demonstrates this for DNNs; linear and logistic regression reduce to feasibility constraints solvable by Z3~\cite{z3}.

\parab{Using \sysname in practice.}
Operators can extend \sysname by adding mechanisms to \cref{sec:phase1} or posing new queries. In our VM allocator study, for instance, \sysname identifies the CPU model as the highest-leverage improvement target---a finding that directly guides retraining investment.

Modeling a new system requires encoding its logic as constraints, which demands familiarity with convex optimization and bi-level reformulation techniques (KKT or primal--dual). This is a one-time cost: once modeled, operators can flexibly explore scenarios, queries, and model variants within the same framework.

The primary contribution of \sysname is diagnostic: it surfaces risks that are invisible to existing analysis methods. Whether operators subsequently deploy guardrails, retrain models, or adjust heuristics is an operational decision---but they cannot act on risks they have not identified. In our evaluation, \sysname revealed that the CPU model dominates migration risk and pinpointed the feature regions responsible; this insight alone justifies the analysis even before any mitigation is deployed.

\parab{LLM-guided search vs.\ formal guarantees.}
LLMs offer a complementary approach: they can propose candidate adversarial patterns quickly by exploiting learned heuristics. However, LLM-generated candidates lack guarantees on optimality or constraint satisfaction---verifying that a proposed scenario respects all statistical and system constraints remains an open challenge~\cite{llm-verify-challenge}. \sysname, by contrast, systematically explores the feasible space and certifies that every reported scenario is reachable under the model's error distribution. A promising hybrid direction is to use LLM-guided proposals as warm starts for \sysname’s optimization, combining speed with formal guarantees.

\parab{Limitations.}
Our evaluation uses three weekdays of production data and the corresponding trained models. \sysname’s slack-based constraints stress-test model behavior within a controlled neighborhood of the observed distribution, providing resilience beyond the literal training window. Nevertheless, longer-term variations---seasonal effects, rare workload spikes, or evolving user behavior---may fall outside this neighborhood. Extending the evaluation to longer horizons and more diverse workloads is a natural next step for the production team.
\section{Related work}
\label{sec:relatedwork}

We distinguish our work from four related areas.


\parab{Empirical analysis.}
Prior work evaluates ML-augmented systems empirically---replaying traces, injecting noise, or sampling perturbations~\cite{xin2021production,nushi2018towards,takanobu2020your}.
Pandora~\cite{nushi2018towards} aggregates historical failures across multiple ML components to quantify each model’s contribution to system-level errors, but depends on past incidents and cannot uncover scenarios that have not yet occurred.
In VM allocation specifically, Wang et al.~\cite{ishailifetime} evaluate a lifetime-aware allocator through trace-driven simulation in Azure, and Ling et al.~\cite{ling2025lava} extend this with distributional lifetime predictions in Google’s clusters. Both demonstrate that ML predictions improve allocator performance, but their analyses remain observational: they assess behavior under sampled workloads, not worst-case combinations.

\sysname differs fundamentally: it systematically searches for realistic worst-case prediction combinations that maximize end-to-end degradation, then maps them back to concrete feature-space conditions. This exposes rare but high-impact risks that empirical evaluation misses due to the combinatorial search space.

\parab{ML safety verifiers.}
Formal verification proves that a model’s input-output behavior satisfies specified properties~\cite{formal-verif-input-output,tornblom2020formal,fischetti2018deep,katz2019marabou,katz2022reluplex,lomuscio2017approach,tjeng2017evaluating}---typically local robustness of a single model invocation.
whiRL~\cite{whirl} extends this to learning-augmented systems, combining neural network verification with system-level transition models to prove safety and liveness of DRL controllers.


These tools focus on correctness, not quantitative performance; they can confirm a model follows its rules but cannot determine how its errors might degrade system throughput or latency. \sysname bridges this gap by quantifying end-to-end performance impact of model interactions and providing concrete scenarios that demonstrate the maximum possible degradation.

\parab{Distributed system analyzers.}
Performance profilers~\cite{fonseca2007x,sigelman2010dapper,kaldor2017canopy,huang2021tprof,aguilera2003performance} diagnose issues in observed executions---e.g., tprof~\cite{huang2021tprof} hierarchically aggregates traces to identify slow operations contributing to tail latency. These tools analyze what \emph{did} happen but do not explore worst-case behavior beyond observed traces, nor reason about systems whose control decisions depend on ML predictions.


Complementary to these observational tools, heuristic analyzers~\cite{metaopt,namyar2024end,fperf,mind-the-gap} encode system heuristics as optimization problems to find adversarial inputs that maximize metrics like latency or resource usage.
However, these analyzers target systems with hand-designed heuristics and deterministic inputs; none can analyze \emph{multi-model} ML-based systems or track how prediction errors from \emph{multiple} probabilistic models propagate through downstream logic.

\parab{Tree-based model analyzers.}
Prior work analyzes perturbation sensitivity of a \textit{single} tree-based model~\cite{chen2019robustness,harden-tree-ensemble,ceagr1,robust-dtree,lp-norm-tree,efficient-adv-attack,robustness-gradient-boost}---an NP-complete problem for general ensembles~\cite{harden-tree-ensemble}. \sysname goes further: it reasons about \emph{multiple} models jointly and traces how their predictions interact with downstream system logic to affect end-to-end performance, not just individual model sensitivity.

\section{Conclusion}
\label{sec:conclusion}



\sysname is the first system that systematically analyzes how multiple ML models jointly affect the performance of a production VM allocator. It models the interactions between CPU, memory, and lifetime predictions and the allocator’s bin-packing heuristic within a principled optimization framework. By capturing both probabilistic model behavior and system dynamics, \sysname uncovers harmful interactions that existing approaches miss. Our evaluation shows that \sysname identifies scenarios with up to $4\times$ higher performance degradation than the current operator's evaluator, which reveals risks that remain hidden under conventional analysis.


\clearpage
\newpage
\bibliographystyle{plain}
\bibliography{reference}

\clearpage
\appendix
\section*{APPENDIX}
\section{\sysname Supported Queries}
\label{sec:sup_queries}
\autoref{table:rc_usecases_extended} summarizes the queries that \sysname supports.

\begin{table*}[h]
\centering
\small

\SetTblrInner{
  rowsep=0pt,
  leftsep=2pt,
  rightsep=2pt
}

\begin{tblr}{
  width=\textwidth,
  colspec={X[0.23,l,m] c X[0.47,l,m] X[0.30,l,m]},
  column{2}={halign=c},
}
\toprule
\textbf{Class of Query} & & \textbf{Query/Question} & \textbf{VM allocator example} \\
\midrule

\SetCell[r=2]{l} \textbf{Worst typical case} &
\SetCell[r=1]{c}\customlabel{circle:1}{\protect\inlinecircle{1}}{\smallcircled{1}} &
How far is the ML-based system from the optimal (over typical workloads)? &
Find the risk of migration or increase in \# utilized servers (\cref{sec:query_1}) \\
\cline{2-4}
& &
What is the impact if we replace or augment a (deployed) algorithm with an ML model or vice versa? &
Can we reduce the risk of migration or \# of servers used? (\cref{sec:query_2}) \\
\midrule

\textbf{The impact of data drift} &
\customlabel{circle:3}{\protect\inlinecircle{3}}{\smallcircled{3}} &
How much data-drift can the ML-augmented system tolerate? &
Can VM allocator tolerate data drift? (\cref{sec:query_3}) \\
\midrule

\SetCell[r=2]{l} \textbf{Identify risk areas} &
&
When and under what conditions does the ML-based system under-perform? &
When should we not use the model's predictions? (\cref{sec:shap}) \\
\cline{2-4}
&
\customlabel{circle:4}{\protect\inlinecircle{4}}{\smallcircled{4}} &
In which area of the ``collective'' feature-space does the system under-perform? Which features contribute to this impact the most? &
Which features to be more careful around? (\cref{sec:shap}) \\
\midrule

\SetCell[r=3]{l}\textbf{Model interactions} &
\customlabel{circle:2}{\protect\inlinecircle{2}}{\smallcircled{2}} &
Which of the deployed ML models causes the worst (typical) case behavior of the system and when? &
Which model should we prioritize to improve? (\cref{sec:query_2}) \\
\cline{2-4}
& &
How should I re-design/replace the algorithms that interact with the ML models to improve worst-case performance?\textsuperscript{*} &
Which VM placement heuristic (best fit, first fit,...) is more resilient to mis-predictions? \\
\cline{2-4}
& &
How do different ML-variants impact the end-to-end performance of the system?\textsuperscript{**} &
What if we replace the models of the VM allocator (LGBMs) with DNNs? \\
\midrule

\textbf{Identify opportunities} &
\customlabel{circle:5}{\protect\inlinecircle{5}}{\smallcircled{5}} &
How much improvement will we get if we improve the accuracy of an ML-model or replace it with another with partially known characteristics? &
How much can we reduce the risk of migration if we improve the CPU model by $10\%$? (\cref{sec:query_9}) \\
\bottomrule
\end{tblr}

\vspace{0.5ex}
\parbox{\textwidth}{\footnotesize
\textsuperscript{*} To answer this query, we simply replace the VM allocator placement heuristic with another (e.g., first-fit, best-fit). It needs to be modeled as a feasibility problem in our heuristic analyzer.\\
\textsuperscript{**} To answer this query, we simply replace SMT solver + CEGAR-based approach in \sysname's second phase with~\cite{namyar2024end}.}

\caption{The types of queries \sysname answers and their mapping to the VM allocator examples. The ``typical'' worst case refers to the worst case under common production workloads. Blue numbered circles correspond to those in \autoref{table:rc_usecases}.}
\label{table:rc_usecases_extended}
\end{table*}

\section{VM Allocator Overview}
\label{sec:vm-allocator}

\begin{figure}[t]
	\centering
	\includegraphics[width=\columnwidth]
	{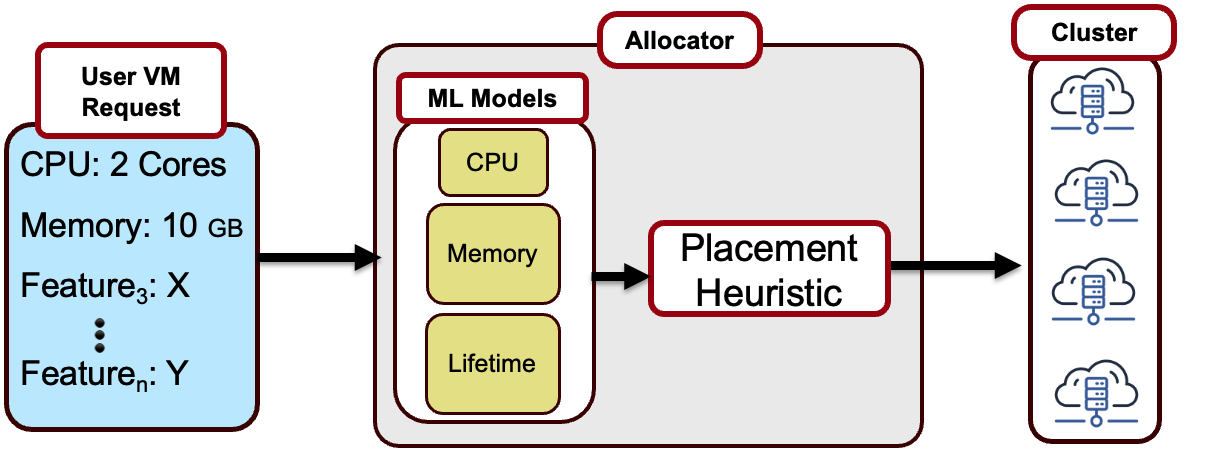}
	\caption{The VM allocator overview.}
	\label{fig:vm-overview}
\end{figure}

The VM allocator (\autoref{fig:vm-overview}) analyzed by \sysname consists of two main components:  
(1) ML models, implemented as Light Gradient Boosting Machines (LGBMs)~\cite{ke2017lightgbm}, that predict VM behavior; and  
(2) a bin-packing heuristic, a variant of best-fit called \emph{Dynamic Preferred Best-Fit Rule} (DPBFR), which uses these predictions to place VMs onto servers (we evaluated the effect of the prototype memory model in \cref{sec:eval}).

\parab{LGBM models.}
Each VM request is represented by a feature vector, including attributes such as VM resource demand (e.g., CPU and Memory), arrival time, and user history. These features serve as inputs to the ML models. The allocator deploys two models~\cite{RC}:
\begin{itemize}
    \item \textbf{CPU utilization model}: predicts whether a VM will have high CPU usage ($\ge 65\%$).
    \item \textbf{Lifetime model}: predicts whether a VM is long-lived (remains in the system for more than 2 hours).
\end{itemize}

The CPU model estimates the number of CPU cores that the VM consumes, while the lifetime model influences which server pool the allocator considers. For memory, the allocator uses the requested amount.

\parab{Placement heuristic.}
The allocator feeds these predictions into the DPBFR heuristic~\cite{ishailifetime} to select a target server. DPBFR assigns each server a \emph{best-fit score} based on the remaining capacity after placing the VM. The score accounts for multiple resource dimensions, such as CPU and memory.

For example, consider a server with 8 CPU cores and 32~GB of memory. If a VM requests 3 CPU cores and 8~GB of memory and the CPU model predicts it will use 2 CPU cores, then placing the VM leaves 6 CPUs and 24~GB of memory available. DPBFR computes normalized remaining-capacity ratios for each resource and aggregates them into a score in $[0,1]$, where lower scores indicate tighter packing.

DPBFR then groups servers into buckets based on their scores. The number of buckets, three or five, depends on whether the lifetime model predicts the VM to be short-lived or long-lived respectively. This design encourages co-locating long-lived VMs, which reduces fragmentation and improves resource utilization. The allocator selects a server uniformly at random from the lowest-score bucket.

\section{Probabilities on Finite Samples}
\label{sec:probsfromsamples}

In \cref{sec:realistic}, we described how the queries that \sysname supports require that we enforce conditions on the fraction of the VMs that have to meet particular constraints. For example, we may want to ensure $p\%$ of predictions across VMs are correct or that $p'\%$ of these predictions are only off by one class. But, as with any statistical analysis on finite samples, we need to translate these conditions into ones that have to hold with high probability (we enforce $99.99\%$ in our case) on any arbitrary finite sample of the distribution we want to enforce.

Our approach is to (1) find the distribution of $p$ (which is an order statistic of the original distribution that we want to enforce) given a finite sample of size $N$; (2) find the range $\mathcal{P} = [ p_s, p_t ]$ where $p \in \mathcal{P}$ with probability $\ge 99.99$ fall within this range; and then (3) add constraints that ensure $y \in \mathcal{P}$ where $y$ is the number of VMs that meet the condition (e.g., where the predictions of the model are correct).

To do this, we use the approach in~\cite{casella2024statistical} which shows that we can use the central limit theorem~\cite{kwak2017central} to approximate the distribution for $p$ as a normal distribution if we have a large enough sample size $N$. Specifically, the rule of thumb is to pick $N$ such that $Np \ge 5 $ and $ N(1-p) \ge 5$. For such a values of $N$, we can assume $p$ has a normal distribution and therefore falls within $p\pm 3.89\sigma_p$ with $99.99\%$ probability according to Chebyshev's inequality~\cite{chebychev} (here $\sigma_p$ is the standard deviation of the normal distribution with the mean equals $p$).

\begin{table*}[t]
\centering
\small
\SetTblrInner{rowsep=1pt}

\begin{tblr}{
  width=\textwidth,
  colspec={X[0.36,l,m] X[0.64,l,m]},
  column{1}={font=\bfseries},
}
\toprule
\textbf{Task}  & \textbf{Technique Summary} \\
\midrule
\textbf{Caching with tree based models~\cite{song2020learning, berger2018towards}} & The user models the caching algorithm in~\sysname as a feasibility problem (this is similar to the packet scheduling model in~\cite{metaopt}~---~we have an example model of caching in~\sysname). They also connect the model's prediction to the weight of the caching model.~\sysname handles the rest.\\
\textbf{Caching with DNNs} & Same as above but replace the CEGAR-based approach with the approach in~\cite{namyar2024end}\\
\textbf{Adaptive bit rate (ABR)~\cite{mao2017neural}} & Need to model the ABR algorithm and the environment it interacts with as a feasibility problem (similar to~\cite{ccac}) and replace the CEGAR-based approach with~\cite{namyar2024end}\\
\bottomrule
\end{tblr}
\caption{\sysname can analyze a diverse set of systems. Here we describe a few such systems and the steps needed to analyze them.}
\label{table:extensibility}
\vspace{-\baselineskip}  
\end{table*}

\section{Modeling LGBMs in an SMT Solver}
\label{sec:lgbm_smt}

Phase~2 searches for a feature vector that makes the deployed models produce a target set of predictions. To do this, we encode each LightGBM model as constraints in an SMT solver. This section describes that encoding and shows how we add cross-model feature constraints when multiple LGBMs share input features.

\subsection{Encoding one LGBM}

An LGBM classifier consists of a set of decision trees. Each tree routes an input feature vector to one leaf and returns that leaf's score. For a multi-class model, LightGBM groups trees by class and sums the scores of all trees in each class. The model then applies a softmax to these class scores and predicts the class with the highest probability.

For our purposes, we do not need to encode the softmax explicitly. Softmax preserves the ordering of class scores, so the predicted class is exactly the class with the maximum total score. We therefore encode only the class-score sums and the corresponding argmax condition.

Let the input feature vector be $\mathbf{f} = (f_1, f_2, \dots, f_m)$. For each tree $t$, we introduce constraints that describe which leaf the input reaches. Each internal node in the tree checks one threshold condition, such as $f_j \le \theta$. A root-to-leaf path is the conjunction of all threshold tests along that path. If the path is satisfied, the tree outputs that leaf's score.

We encode each tree with:
\begin{itemize}
    \item one Boolean or arithmetic condition for each branch test,
    \item one variable for the tree output score, and
    \item constraints that tie the score to exactly one feasible leaf.
\end{itemize}

For each class $c$, we sum the outputs of its trees:
\begin{align*}
    S_c = \sum_{t \in \mathcal{T}_c} \text{score}_t.
\end{align*}

To force the model to predict a target class $c^\star$, we add:
\begin{align*}
S_{c^\star} \ge S_c \qquad \forall c \neq c^\star.
\end{align*}

This is enough to encode the prediction outcome.

\subsection{A small example}

Consider a binary LGBM with one tree for class~0 and one tree for class~1. The model uses two features, $f_1$ and $f_2$.

The tree for class~0 is:
\begin{align*}
    &\text{if } f_1 \le 4 \text{ then score } = 0.3,\\
    &\text{else if } f_2 \le 2 \text{ then score } = 0.8,\\
    &\text{else score } = 0.1. 
\end{align*}

The tree for class~1 is:
\begin{align*}
&\text{if } f_2 \le 5 \text{ then score } = 0.2,\\
&\text{else score } = 0.9.
\end{align*}

We introduce score variables $s_0$ and $s_1$ for the two trees. We encode
\begin{align*}
 &s_0 \in \{0.3, 0.8, 0.1\}, \\
&s_1 \in \{0.2, 0.9\},   
\end{align*}
with the constraints
\begin{align*}
&f_1 \le 4 \Rightarrow s_0 = 0.3, \\
&f_1 > 4 \wedge f_2 \le 2 \Rightarrow s_0 = 0.8,\\
&f_1 > 4 \wedge f_2 > 2 \Rightarrow s_0 = 0.1,
\end{align*}
and
\begin{align*}
&f_2 \le 5 \Rightarrow s_1 = 0.2, \\
&f_2 > 5 \Rightarrow s_1 = 0.9.
\end{align*}
Since there is one tree per class, the class scores are $S_0=s_0$ and $S_1=s_1$.
To force prediction of class~1, we add $S_1 \ge S_0$.

Now consider the feature vector $(f_1, f_2) = (6, 6)$. The class~0 tree returns $0.1$, the class~1 tree returns $0.9$, and the model predicts class~1 since $0.9 \ge 0.1$.

This example shows the core idea: each tree contributes a score, and the SMT solver searches for feature values that make the target class score dominate.

\subsection{Adding multiple LGBMs}

The VM allocator uses more than one model. For example, the CPU model and the lifetime model both take a VM description as input, and some features appear in both models. In Phase~2, we must find one feature vector that makes all models produce their target predictions at the same time.

We handle this by encoding all LGBMs in the same SMT query and reusing the same SMT variables for shared features.

Suppose the CPU model uses features
\[
(f_{\text{vmtype}}, f_{\text{hour}}, f_{\text{histCPU}})
\]
and the lifetime model uses
\[
(f_{\text{vmtype}}, f_{\text{hour}}, f_{\text{histLife}}).
\]

The two models share $f_{\text{vmtype}}$ and $f_{\text{hour}}$. We create one SMT variable for each shared feature:
\[
x_{\text{vmtype}},\quad x_{\text{hour}}.
\]

We then use these same variables in both model encodings. This forces both LGBMs to reason about the same VM. The remaining model-specific features get their own variables:
\[
x_{\text{histCPU}},\quad x_{\text{histLife}}.
\]

If Phase~1 says that a VM should produce a target CPU prediction and a target lifetime prediction, we add both prediction constraints to the same solver instance. The SMT solver then searches for one assignment
\[
(x_{\text{vmtype}}, x_{\text{hour}}, x_{\text{histCPU}}, x_{\text{histLife}})
\]
that satisfies both models simultaneously.

\subsection{Feature constraints across models}

Multiple models often impose additional consistency requirements beyond shared input variables. We can express these requirements directly as SMT constraints.

\parab{Shared categorical features.}
If both models use the same VM type, we use one variable for that feature in both encodings. This already enforces consistency.

\parab{Shared numerical features.}
If both models use the same arrival hour or VM size, we again reuse one variable. The solver cannot choose different values for different models.

\parab{Domain constraints.}
We can restrict features to valid ranges:
\[
1 \le x_{\text{hour}} \le 24,\qquad
x_{\text{memory}} \in \{2,4,8,16,32\}.
\]

\parab{Cross-feature constraints.}
We can also encode relations between features. For example, if a large VM type always implies at least 16~GB of requested memory, we add:
\[
x_{\text{vmtype}} = \texttt{large} \Rightarrow x_{\text{memory}} \ge 16.
\]

These constraints help the solver avoid infeasible or unrealistic feature assignments.

\subsection{Why this matters for \sysname}

This SMT encoding lets Phase~2 do more than check one model at a time. It lets us ask a joint question: \emph{Does there exist a single feature vector that makes all deployed models produce the target predictions found in Phase~1?}

If the answer is yes, the solver returns such a feature vector. If the answer is no, the target prediction pattern is infeasible at the feature level. This is exactly why the multi-model SMT query matters: it rules out combinations of predictions that look valid in isolation but cannot arise from any real VM description.

In practice, large LGBMs make this encoding expensive. This is why \sysname adds the CEGAR-based abstraction--refinement loop described in \cref{sec:CEGAR}. The SMT formulation here is the base model. CEGAR-based approach makes it scale.

\section{Joint Prediction and Ground-Truth Mapping with a CEGAR-Based Approach}
\label{sec:ground-truth-cegar}

We extend our CEGAR-based approach to jointly recover a feature vector $\mathbf{f}$ that produces both (i) the target predictions of the deployed LGBMs and (ii) the corresponding ground-truth labels that Phase~1 identified. 

To achieve this, we need a mechanism that maps features to their ground-truth labels, i.e., a function $\mathcal{F}(\mathbf{f})$. However, such a mapping is not available in closed form, and modeling it as a discrete lookup table does not scale. Instead, we leverage production data, which provides samples of feature vectors paired with their ground-truth labels, giving us a partial view of $\mathcal{F}(\mathbf{f})$.

We approximate this mapping using an \emph{overfit reference model}. Specifically, we train a model that achieves near-perfect accuracy on the available production data and use it as a proxy for $\mathcal{F}(\mathbf{f})$. While such models generalize poorly and are unsuitable for prediction, they effectively memorize the observed data and serve as a scalable approximation of a lookup map.

We train a reference LGBM for each model (e.g., CPU and lifetime) and incorporate these reference models into the CEGAR-based framework. The SMT solver then searches for a feature vector $\mathbf{f}$ such that:
(i) the deployed models produce the target predictions identified in Phase~1, and 
(ii) the reference models produce the corresponding ground-truth labels.

This joint formulation enables Phase~2 to recover feature vectors that are consistent with both the desired prediction behavior and the observed ground-truth patterns in production data.

\section{How to Apply~\sysname to Other Problems}
\label{sec:extensibility}
In this paper we focused on how to use~\sysname to analyze a VM allocator system in production. But~\sysname applies to a wide range of problems. We discuss a few examples and how users can extend~\sysname to those examples in~\autoref{table:extensibility}. 

For each new system a user wants to analyze, similar to MetaOpt~\cite{metaopt}, they would have to first model their system either as a feasibility probelm or a convex optimization. Once we have this model,~\sysname uses it to produce the answers to queries in~\autoref{table:rc_usecases}.

\section{The Overall Risk Surface for a VM Sequence}
\label{sec:risk_surface}

We showed that how one can compute a risk surface for the model when we analyze systems where a single prediction of each model determines the system's end-to-end performance. But in systems like the VM allocator we analyze a sequence of predictions and need to merge these risk surfaces (which show the risk surface for each VM in the sequence) to find the overall risk surface.

Our goal is to find the area in the ``adversarial feature space'' (i.e., what we find as the risk surface for each VM in the sequence) which is the most common across all VMs in the sequence. Our approach is simple, we score each region of the feature space based on the number of individual risk-surfaces it appears in. This means the more VMs that share that region of the feature space in their adversarial feature space, the higher the score of that region is in our analysis. We then return the areas in the feature space with the highest score as the overall risk surface of the sequence.

The entropy for the values of a given feature in its risk surface (compared to its full range) determine whether that feature plays a critical role in the performance of the system or not. For example, in our analysis of the VM allocator we see that multiple features have a risk surface with maximum possible entropy (i.e., the value of the feature in its risk surface spans the entire range of values that feature can take): these features do not pose a critical threat to the system.

One can imagine different variants of this approach: we can have a cut-off for the minimum number of risk surfaces that a sub-region needs to belong to before it qualifies in the sequence adversarial space or we can have a strict requirement where we only consider a sub-region as part of the risk surface if it appears in every VM's adversarial feature space. Our goal here is to show the potential of~\sysname to facilitate such analysis and the study of which of these metrics is the right one to use is beyond the scope of this work.

\section{How We Model LGBMs in MetaOpt}
\label{sec:LGBM}

We next show how to model a light gradient-boosting machine (LGBM) in MetaOpt. MetaOpt can analyze a heuristic if we model it as a convex optimization or a feasibility problem which a solver like Guroubi~\cite{gurobi} can handle (these include mixed integer problems, bilinear problems, \dots). We model LGBMs as a feasibility problem and encode it as mixed-integer linear problem.

We separate the stages involved in running inference for an LGBM: (1) each decision tree in the ensemble produces a value (the score $S$) based on the feature vector; (2) the LGBM combines these scores for each class (into a vector $P'$) and then computes a softmax over them to produce a probability distribution (for each class $i$ it produces a number $0 \le P_i \le 1$ which is the probability with which sample $i$ belongs to class $i$); (3) it then computes the maximum likelihood estimate given this probability distribution. 

We observe that for the purpose of computing the predicted class label we can skip the softmax computation:

\begin{align*}
    i = \text{argmax}_{i} (P = \text{softmax}(P')) = \text{argmax}_{i} P'
\end{align*}

This observation allows us to avoid non-linear constraints within our model. All we need now is to find the class with the maximum score, $i =  \text{argmax}_{i} P'$. This means that we need to have an indicator variable (one that can only take the values $0$ or $1$) which shows which index $i$ corresponds to the $P'_i$ which is the highest among all $P'$. We use the binary variable $x_i$ and enforce constraints that only allows $x_i = 1$ iff $P'_i \ge P'_j$ for all values of $j$. 
These constraints are:

\begin{align}
& P'_j - P'_i \le M (1 - x_i) \quad \forall i,j  \mid i \neq j \nonumber \\
&\sum x_i = 1 \label{eq:maxLikelihood}
\end{align}

\begin{figure}[t]
    \centering
    \begin{tikzpicture}[
    every node/.style={font=\sffamily},
    leaf/.style={fill=teal!15,draw=black,line width=0.3pt,minimum width=0.6cm,minimum height=0.6cm,rounded corners=2pt,inner sep=0pt},
    thickgray/.style={thick,gray!70}
  ]
\def\squaresep{0.7} 
\def\leafsize{0.6}  
\def\loffset{0.5*\leafsize}

\node[draw,fill=red!20,circle,minimum size=1cm] (root) at (0,0) {};

\node[draw,fill=red!20,circle,minimum size=1cm] (left) at (-2.2,-2) {};
\node[draw,fill=red!20,circle,minimum size=1cm] (right) at (2.2,-2) {};

\draw[thick] (root) -- (left);
\draw[thick] (root) -- (right);

\node[font=\itshape,scale=1.2] at (-2.2,-0.6) {$f_{\text{feature(k)}} \leq T_{kj}$};
\node[font=\itshape,scale=1.2] at (2.2,-0.6) {$f_{\text{feature(k)}} > T_{kj}$};

\foreach \i in {0,...,4} {
  \node[leaf] (l\i) at ({-4+\i*\squaresep},-3.6) {};
}
\foreach \i in {0,...,4} {
  \node[leaf] (r\i) at ({0.6+\i*\squaresep},-3.6) {};
}

\coordinate (lstart) at ($(-4-\loffset,-3.6)$);
\coordinate (lend)   at ($(-4+4*\squaresep+\loffset,-3.6)$);
\coordinate (rstart) at ($(0.6-\loffset,-3.6)$);
\coordinate (rend)   at ($(0.6+4*\squaresep+\loffset,-3.6)$);

\draw[thick] (left) -- ($(-4-\loffset,-3.6) + (0,0.3)$);
\draw[thick] (left) -- ($(-4+4*\squaresep+\loffset,-3.6) + (0,0.3)$);

\draw[thick] (right) -- ($(0.6-\loffset,-3.6) + (0,0.3)$);
\draw[thick] (right) -- ($(0.6+4*\squaresep+\loffset,-3.6) + (0,0.3)$);

\draw[thick, decorate, decoration={brace, mirror, amplitude=10pt}] 
  ($(-4-\loffset,-3.6) + (0,-0.4)$) -- 
  ($(-4+4*\squaresep+\loffset,-3.6) + (0,-0.4)$);
\node[anchor=north, font=\itshape, scale=1.4] at ($(-4-\loffset,-3.6)!0.5!(-4+4*\squaresep+\loffset,-3.6) + (0,-0.9)$) {$\mathcal{L}_{kj}$};

\draw[thick, decorate, decoration={brace, mirror, amplitude=10pt}] 
  ($(0.6-\loffset,-3.6) + (0,-0.4)$) -- 
  ($(0.6+4*\squaresep+\loffset,-3.6) + (0,-0.4)$);
\node[anchor=north, font=\itshape, scale=1.4] at ($(0.6-\loffset,-3.6)!0.5!(0.6+4*\squaresep+\loffset,-3.6) + (0,-0.9)$) {$\mathcal{R}_{kj}$};

\end{tikzpicture}
\caption{LGBM modeling in MetaOpt}
\label{fig:decision-tree-in-metaopt}
\end{figure}

\begin{figure}
\centering
\begin{tikzpicture}
\begin{axis}[
    width=0.9\linewidth,
    height=0.5\linewidth,
    xlabel={Number of Trees},
    ylabel={Runtime (seconds)},
    xmode=log,
    log basis x={10},
    xtick={1,10,100,1000},
    minor x tick num=9,
    grid=both,
    grid style={dashed, gray!30},
    mark size=2pt,
    thick,
    ymajorgrids=true,
    grid=both,
    ymajorgrids=true,
    yminorgrids=false,     
    xtick align=outside,
    xtick pos=bottom,
    x tick style={      
        color=black,      
        line width=0.6pt,
    },
    y tick style={      
        color=black,      
        line width=0.6pt,
    },
    ytick pos=left,
    major tick length=1mm,
    minor y tick num=4, 
    minor y tick style={color=gray, line width=0.4pt},
    minor tick length=0.75mm,
    grid style={dashed, lightgray, dash pattern=on 1pt off 1pt},
]

\addplot+[mark=*, color=blue] coordinates {
    (2, 4)
    (30, 25)
    (250, 191)
    (500, 394)
    (1000, 734)
};

\end{axis}
\end{tikzpicture}
\caption{Runtime vs. number of trees.}
\label{fig:lgbm-runtime}
\end{figure}

Here $M$ is the traditional ``Big-M'' method (a large number). The second sum in~\autoref{eq:maxLikelihood} ensures we only return one class as the output of the LGBM (the optimization is free to break ties in any way it chooses).


We next need to find $P'_i$. We know $P'_i = \sum_{j=1}^{n_i} S^i_j ~\forall i$, where $n_i$ is the number of decision trees for class $i$, and $S^i_j$ is the score of tree $j$ of class $i$. To compute this sum, we first need to find $S^i_j$~---~we need to find which leaves in each tree are ``activated'' and return the value that the tree returns at those leaves.

Luckily the structure of the tree allows us to do this efficiently: For each class $i$, the $j^{\text{th}}$ tree applies a threshold $T_{kj}$ at each node $k$ to a feature $f_{\text{feature}(k)}$ (the function $\text{feature}(k)$ finds which feature the node $k$ acts on) and takes the left branch if the feature is less than the threshold and the right one otherwise (\autoref{fig:decision-tree-in-metaopt}). At each node we restrict the set of leaves that are allowed to be active based on the value of the feature at that node. The set of all such constraints allows us to find $S_j$. 

Formally, we use the variable $a_{lj}$ as an indicator of whether the $l^{\text{th}}$ leaf in tree $j$ is active. We have:

\begin{align*}
& \sum_{l} a_{lj} = 1 \\
& \sum_{l \in \mathcal{L}_{kj}} a_{lj} \le 1 + \epsilon (T_{kj} - f_{\text{feature}(k)}) \quad \forall k \\
& \sum_{l \in \mathcal{R}_{kj}} a_{lj} \le 1 + \epsilon (f_{\text{feature(k)}} - T_{kj}) - \mu \quad \forall k
\end{align*}

where $\mathcal{L}_{kj}$ and $\mathcal{R}_{kj}$ are all the leaf nodes to the left and to the right of the $k^{th}$ node in the $j^{th}$ tree. The two values $\epsilon$ and $\mu$ are small numbers. These constraints ensure that only one leaf is activated per tree. We then simply assign the value of the activated leaf to $S^i_j$.

This set of constraints fully models the inference behavior of an LGBM. Additionally, \autoref{fig:lgbm-runtime} illustrates how the runtime of MetaOpt increases as the number of trees in the LGBM model grows, when predicting the output for 1000 feature vectors. The observed non-linear growth in runtime indicates that larger models require more computational time and may be harder to scale efficiently.



\begin{figure}[t]
    \centering
    \resizebox{0.95\columnwidth}{!}{
\definecolor{mone}{RGB}{205,230,245}   
\definecolor{mtwo}{RGB}{245,215,225}   
\definecolor{resu}{RGB}{210,235,200}   

\tikzset{
  circnode/.style = {circle, minimum size=22pt, inner sep=0pt, thick, draw=black},
  Mone/.style     = {circnode, fill=mone},
  Mtwo/.style     = {circnode, fill=mtwo},
  Rnode/.style    = {circnode, fill=resu},
  eb/.style       = {line width=1pt, draw=black},             
  er/.style       = {line width=2pt, draw=red},               
  ed/.style       = {line width=1.5pt, draw=magenta!70, dashed}, 
  pathtxt/.style  = {text=black, font=\bfseries\footnotesize}
}

\begin{tikzpicture}[x=2.4cm,y=2.0cm, font=\boldmath]

\node[Mone] (m10) at (0, 1) {$\mathcal{M}_1^{0}$};
\node[Mone] (m11) at (0, 0) {$\mathcal{M}_1^{1}$};
\node[Mone] (m1m) at (0,-1) {$\mathcal{M}_1^{-1}$};

\node[Mtwo] (m20) at (1.7, 1) {$\mathcal{M}_2^{0}$};
\node[Mtwo] (m21) at (1.7, 0) {$\mathcal{M}_2^{1}$};
\node[Mtwo] (m2m) at (1.7,-1) {$\mathcal{M}_2^{-1}$};

\node[Rnode] (r0) at (3.5, 1) {$R^{0}$};
\node[Rnode] (r1) at (3.5, 0) {$R^{1}$};
\node[Rnode] (rm) at (3.5,-1) {$R^{-1}$};

\draw[eb] (m10) -- (m20);         
\draw[eb] (m10) -- (m21);         
\draw[eb] (m10) -- (m2m);         
\draw[er] (m11) -- (m20);         
\draw[er] (m11) -- (m21);         
\draw[ed] (m1m) -- (m20);         
\draw[ed] (m1m) -- (m2m);         

\draw[eb] (m20) -- (r0);          
\draw[eb] (m21) -- (r1);          
\draw[eb] (m2m) -- (rm);          
\draw[er] (m21) -- (r0);          
\draw[ed] (m2m) -- (r0);          
\draw[ed] (m20) -- (r1);          
\draw[er] (m20) -- (rm);          

\node[pathtxt, rotate=0 ] at ($(m10)!0.35!(m20)+(-0.15,0.10)$) {Path1};
\node[pathtxt, rotate=335] at ($(m10)!0.38!(m21)+(-0.18,0.20)$) {Path2};
\node[pathtxt, rotate=315] at ($(m11)!0.38!(m20)+(-0.18,0.20)$) {Path3};


\end{tikzpicture}}
     \caption{Flow-graph of two models (\mone, \mtwo) and their relative difference $R= $\mtwo$-$\mone. Each node $M_x^y$ denotes model $x$ predicting with offset $y\in\{0,1,-1\}$ (0=correct, 1=over, -1=under) compared to the ground truth. Nodes $R^y$ show output differences. Paths (e.g., Path1–3) are feasible joint outcomes, with flow values encoding the fraction of VMs. This enforces both individual accuracy and relative-output constraints.}
    \label{fig:flowgraph}
\end{figure}

\section{Implementation Details of Hypothetical Model} 
\label{sec:flow_model_hypothetical}
In query~\ref{query:5}, the operator wants to estimate the performance gain from replacing the current model (\mone), e.g. the CPU model,  with a more accurate hypothetical one (\mtwo) (We describe our approach for a \emph{binary} model to match the model deployed in production~\cite{ishailifetime}, but the approach also extends to \emph{multi-class} models). The current model is $x_1\%$ accurate, overpredicts $x_2\%$, and underpredicts $x_3\%$. The goal is a model that is $x_1+10\%$ accurate with less $5\%$ over- or underprediction.

The hypothetical model (\mtwo) must also match \mone's output $y_1\%$ of the time, output ``True'' when \mone outputs ``False'' $y_2\%$ of the time, and vice versa $y_3\%$ of the time. These constraints reflect operational realities. In production, new models are often deployed gradually (e.g., A/B testing or shadow deployment) alongside the existing model, and large discrepancies in predictions can trigger unexpected scheduling or placement behaviors that operators are not prepared for. Keeping a high match rate ensures the new model can be adopted without destabilizing downstream heuristics, capacity planning assumptions, or SLAs. The mismatch cases ($y_2\%$, $y_3\%$) quantify acceptable divergence—enough to capture improvements without introducing untested behavior into most allocation decisions.

If the models were independent, we could encode these constraints separately using mechanisms 1 and 2 from \cref{sec:realistic}. But they are dependent: the relative-output fractions (\mtwo $-$ \mone) depend on both models’ predictions for each VM. Here, True is mapped to $1$ and False to $0$, so a match gives \mtwo $-$ \mone = 0, \mtwo overpredicting gives 1, and underpredicting gives -1. These relative differences must satisfy the proportion constraints of $y_1\%$, $y_2\%$, and $y_3\%$ across the entire sequence. This means the accuracy of each model is tied to their joint output distribution.

Modeling both accuracy targets and relative-output dependencies together makes the problem non-trivial. This challenge led us to design a novel graph-based approach

The relative difference between the models' outputs is defined as  
$R =$ \mtwo $-$ \mone $\in \{0, 1, -1\}$.  
Here, $R = 0$ means \mone and \mtwo agree, $R = 1$ means \mtwo outputs True while \mone outputs False, and $R = -1$ means \mtwo outputs False while \mone outputs True.  

\autoref{fig:flowgraph} shows three sets of nodes, one for \mone, one for \mtwo, and one for $R$. A node $\mathcal{M}_x^y$ denotes model $x$ with a difference $y$ from the ground truth where $y \in \{0,1,-1\}$. A node $R^y$ denotes that the relative difference between \mone and \mtwo is $y$.  
Each \emph{path} from an \mone node to an \mtwo node to an $R$ node corresponds to a feasible joint prediction (\mone, \mtwo, $R$) for a VM. For example:  
\begin{itemize}
    \item \textbf{Path1:} $\mathcal{M}_1^0 \rightarrow \mathcal{M}_2^0 \rightarrow R^0$ — both models match the ground truth and agree.  
    \item \textbf{Path2:} $\mathcal{M}_1^0 \rightarrow \mathcal{M}_2^1 \rightarrow R^1$ — \mone matches the ground truth, \mtwo overpredicts (predicts True when ground truth is False).  
    \item \textbf{Path3:} $\mathcal{M}_1^0 \rightarrow \mathcal{M}_2^{-1} \rightarrow R^{-1}$ — \mone matches the ground truth, \mtwo underpredicts (predicts False when ground truth is True).  
\end{itemize}

We assign each path a \emph{flow}, $P(\texttt{Path}_j)$, which is the fraction of all VMs in the sequence that exhibit that exact joint outcome. The flow is constant along the entire path. For example, every edge in Path2 carries the same $P(\texttt{Path}_2)$, and If $P(\texttt{Path}_2)$ = 0.05, then exactly 5\% of all VMs follow that models output pattern.

A node’s \emph{capacity} $C(\texttt{Node}_i)$ is the total fraction of VMs that exhibit the property represented by that node (e.g., ``\mone predicts correctly'' or ``\mtwo overpredicts''). This fraction comes from the marginal accuracy or error rates that we want to enforce for that model. For example, if \mone is $70\%$ accurate, then $C(\texttt{M}_1^0) = 0.7$.

Because the predictions of \mone and \mtwo for each VM follow exactly one path, the capacity of a node is equal to the sum of the flows of all paths that pass through it:
\begin{align*}
    C(\texttt{Node}_i) = \sum_{j \in \text{Path indices that go through node } i} P(\texttt{Path}_j)
\end{align*}
For example:  
\begin{align*}
    C(\mathcal{M}_1^0) = P(\texttt{Path}_1) + P(\texttt{Path}_2) + P(\texttt{Path}_3)    
\end{align*}

This creates a natural \emph{flow conservation} rule:  
\begin{itemize}
    \item Each node’s incoming flow equals its outgoing flow.
    \item The Node capacity comes from the target accuracy or error rate for that model.
    \item Paths are mutually exclusive — no VM can take two paths.
\end{itemize}

Therefore, this problem converts to finding a feasible assignment of path flows that satisfies all node capacities simultaneously, which we solve using an linear programming (LP) formulation.

For the scenario described earlier, we set:  
\begin{align*}
  C(\mathcal{M}_1^0) = 0.70, \quad C(\mathcal{M}_1^1) = 0.15, \quad C(\mathcal{M}_1^{-1}) = 0.15 \\
  C(\mathcal{M}_2^0) = 0.80, \quad C(\mathcal{M}_2^1) = 0.10, \quad C(\mathcal{M}_2^{-1}) = 0.10 \\
  C(R^0) = 0.90, \quad C(R^1) = 0.05, \quad C(R^{-1}) = 0.05
\end{align*}

We feed this graph into an LP solver (Gurobi~\cite{gurobi}). If no feasible set of path flows exists, the hypothetical model’s targets are incompatible with the current model. If a feasible solution is found, the solver outputs $P(\texttt{Path}_j)$ for all paths, and we enforce these fractions as constraints in MetaOpt. 

For example, if $P(\texttt{Path}_1) = 0.6$, this means that both models match the ground truth for 60\% of the VMs in the sequence. We randomly select 60\% of the VMs and enforce constraints that force the outputs of \mone and \mtwo to match the ground truth (see \autoref{fig:mechanisms_examples}). We implement this by jointly constraining the binary variables of both models for each selected VM to indicate whether the prediction matches the ground truth. We repeat this process for each joint distribution and their corresponding binary variables.


\section{Implementation of VM Placement Algorithm and System's Performance Metric}
\label{sec:dpbfr}





\subsection{Modeling the placement heuristic}

\cref{sec:vm-allocator} explained how the placement algorithm, Dynamic Preferred Best-Fit Rule (DPBFR)~\cite{ishailifetime}, operates in the VM allocator.
We use binary variables to apply the quantization, and we approximate the random (uniform) bin selection of DPBFR with another set of binary variables. We can also enforce round-robin bin selection from the set of candidate servers: this mimics DPBFR's performance in the average case. To model the worst-case operators, we can remove this constraint, which then allows MetaOpt to select the worst-case selection of servers.

We formulate DPBFR as a feasibility problem for the VM allocator in both $H(I)$ and $H'(I')$ in \autoref{eq:bi-level}, defined by a set of constraints without an explicit objective. The notation used is summarized in \autoref{tab:dpbfr-notations}.

\begin{table}[t]
\centering
\small
\setlength{\tabcolsep}{4pt} 
\renewcommand{\arraystretch}{1.05} 

\begin{tabular}{l p{0.85\columnwidth}}
\toprule
\textbf{Term} & \textbf{Meaning} \\
\midrule
$i,j,d$ & indexes for VM, server and dimension \\
$Y_i^d$ & Predicted VM $i$ size in dimension $d$, obtained from either the ML model or ground-truth predictions \\
$\mathcal{Y}_i^d$ & Actual VM $i$ size in dimension $d$ \\
$C_j^d$ & Server $j$ capacity in dimension $d$ \\
$life_i$ & Lifetime of VM $i$ \\
$a_i$ & Arrival time of VM $i$ \\
$l_i$ & Prediction of lifetime model for VM $i$ \\
$\alpha_{ijt}$ & $= 1$ if VM $i$ is on server $j$ at time $t$ (0 otherwise) \\
$\alpha'_{ij}$ & $= 1$ if server $j$ is a candidate for VM $i$ (0 otherwise) \\
$x_{ijt}^d$ & VM $i$ occupied space on server $j$ in dimension $d$ at time $t$ \\
$score_{ij}$ & Best-fit score of VM $i$ for server $j$ \\
$f_{ij}$ & $= 1$ if ball $i$ can fit in bin $j$ (0 otherwise) \\
\bottomrule
\end{tabular}

\caption{Notation of placement heuristic and system's performance metric formulation.}
\label{tab:dpbfr-notations}
\end{table}

We define these constraints step by step:

\parab{Modeling VM placements in servers.} 
VMs arrive over a defined time horizon $T$ (e.g., one day), and DPBFR places each incoming $\text{VM}_i$ starting at its arrival time $a_i$, keeping it active until its lifetime ends at $a_i + life_i$. Let $\alpha_{ijt}$ be a binary variable denoting whether $\text{VM}_i$ is assigned to $\text{server}_j$ at time $t$. Suppose $J$ is the total number of servers. The placement constraints are defined as follows:
\begin{align}
& \alpha_{ijt} = 0 \quad \forall{i,j}, \forall{t} \le a_i \label{eq:dpbfr-11} \\
& \sum_{j=1}^{J}\alpha_{ija_i} = 1 \quad \forall{i} \label{eq:dpbfr-12} \\
& \alpha_{ijt+1} \le \alpha_{ijt} \quad \forall{i, j, t \ge a_i} \label{eq:dpbfr-13} \\
& life_i \le \sum_{j=1}^{J}\sum_{t=a_i}^{T} \alpha_{ijt} < life_i + 1 \quad \forall{i} \label{eq:dpbfr-14}
\end{align}

\autoref{eq:dpbfr-11} ensures that no VM is assigned before its arrival time. \autoref{eq:dpbfr-12} guarantees that each VM is initially placed on exactly one server. \autoref{eq:dpbfr-13} enforces continuity—once a VM is assigned to a server, it cannot leave and return later. Finally, \autoref{eq:dpbfr-14} ensures that the VM stays allocated for its lifetime.

\parab{Modeling capacity constraints and best-fit score.}
We first make sure that $\text{VM}_i$ is assigned to a server that has enough capacity at VM's arrival. Let $x_{ijt}^d$ denote the occupied space of $\text{VM}_i$ in $\text{server}_j$ for dimension $d$ (CPU or Memory), and $Y_i^d$ is the total size of $\text{VM}_i$ in dimension $d$. We have $Y_i^d \alpha_{ijt} = x_{ijt}^d$, and to linearize this equation, we use a technique similar to big-M approach in optimization literature. Let $M$ be a sufficiently large positive integer constant:
\begin{align}
    & x_{ijt}^d \le M \alpha_{ijt} \quad \forall{i, j, t, d}\label{eq:dpbfr:21} \\
    & \sum_{j=1}^{J} x_{ijt}^d \le Y_i^d \quad \forall{i, t, d} \label{eq:dpbfr:22} \\
    & \sum_{j=1}^{J} x_{ijt}^d \ge Y_i^d - M(1-\sum_j \alpha_{ijt}) \quad \forall{i, t, d} \label{eq:dpbfr23} 
\end{align}

Then we define $r_{ij}^d$ as the remained capacity in dimension $d$ of $\text{server}_j$ if $\text{VM}_i$ is placed at its arrival time $a_i$. Let $C_j^d$ be the $\text{server}_j$'s total capacity: 
\begin{align}
    & r_{ij}^d = C_j^d - Y_i^d - \sum_{\text{VM }u < i} x_{uja_i}^d \quad \forall{i, j, d}
\end{align}

The sum on the right captures the residual capacity of $\text{server}_j$ in dimension $d$, and we define a binary variable $f_{ij}$ that shows whether $\text{VM}_i$ can fit in $\text{server}_j$ in all the dimensions or not. The following constraint captures this logic:
\begin{align}
    \min r_{ij}^d \le Mf_{ij} \le M + \min r_{ij}^d 
\end{align}

Finally we define the best-fit score of $\text{VM}_i$ for $\text{server}_j$ as $score_{ij} \in [0,1]$:
\begin{align}
    score_{ij} = \sum_d \frac{C_{j}^d - Y_i^d - \sum_{u < i}x_{uja_i}^d}{C_j^d} \quad \forall{i, j}
\end{align}


\parab{Quantizing the best-fit score.}  
We quantize the best-fit score, ($score_{ij}$) based on the predicted lifetime label \( l_i \) for \( \text{VM}_i \), as given by the lifetime model. If \( l_i = 0 \), indicating a short-lived VM, we divide the score range into 3 buckets, for example: \(\{[lb_1, ub_1), [ub_1, ub_2), [ub_2, ub_3]]\}\). If \( l_i = 1 \), representing a long-lived VM, we use 5 quantization buckets instead.

Let \( b_{ij} \) denote the bucket index that \( score_{ij} \) falls into for server \( j \). For instance, if \( l_i = 0 \) and \( score_{ij} \in [lb_1, ub_1) \), then \( b_{ij} = 1 \). The full set of bucket assignment rules is based on combinations of \( l_i \) and \( score_{ij} \), and each condition can be directly expressed using linear constraints in an LP formulation.

\parab{Determining the servers with lowest bucket index.}  
To select a placement for $\text{VM}_i$, we focus on servers with the lowest bucket index values, as they correspond to the best (i.e., tightest) fits. Among these candidates, one server will be chosen randomly. We define $z_i$ as the minimum bucket index across all servers for $\text{VM}_i$, and introduce a binary variable $\alpha'_{ij}$ that is set to 1 if and only if the bucket index $b_{ij}$ of $\text{server}_j$ equals $z_i$. The following constraints enforce this logic:
\begin{align}
    & z_i \le b_{ij} + M(1 - f_{ij}) \quad \forall{i,j} \label{eq:dpbfr31} \\
    & \alpha'_{ij} \le f_{ij} \quad \forall{i,j} \label{eq:dpbfr:32} \\
    & \alpha'_{ij} \le 1 + \epsilon(z_i - b_{ij}) \quad \forall{i,j} \label{eq:dpbfr:33} \\
    & \alpha'_{ij} \ge -M(1 - f_{ij}) + \epsilon(z_i - b_{ij}) + \epsilon \quad \forall{i,j} \label{eq:dpbfr:34}
\end{align}


\parab{Random selection of candidate servers.}  
To randomly select one of the candidate servers for placing $\text{VM}_i$, we generate a list $Rand$ containing a random permutation of server indices from 1 to the total number of servers $J$. We then adopt the same technique used in \cite{metaopt} for modeling the first-fit heuristic, but modify it to pick the first server in $Rand$ for which $\alpha'_{ij} = 1$—that is, the first eligible candidate server for $\text{VM}_i$. The following constraint enforces this random selection:
\begin{align}
    & \alpha_{ija_i} \le \frac{\alpha'_{iRand[j]} + \sum_{\text{server}_{k<j}}(1 - \alpha'_{iRand[k]})}{j} \quad \forall{i, j}
\end{align}

\subsection{Modeling the system performance metric}
\label{sec:objective-formulation}

After encoding the placement heuristic, we model the system performance metric ($H(\cdot)$ or $H'(\cdot)$ in \autoref{eq:bi-level}). In our setting, this metric captures either (i) the number of overutilized servers or (ii) the number of active (utilized) servers at each timestep $t$. We describe both below.

\parab{Server overutilization.}
We mark a server as overutilized if its demand exceeds capacity in any resource dimension. Specifically, server $j$ is overutilized in dimension $d$ at time $t$ if:
\begin{align}
    \texttt{IF } \bigl(C^d_{jt} < \sum_i \mathcal{Y}^d_i \cdot \alpha_{ijt}\bigr)\ \texttt{THEN }\; \textit{overutil}^d_{jt} = 1 \label{eq:overutil-jtd}
\end{align}

Here, $\textit{overutil}^d_{jt}$ is a binary variable indicating whether server $j$ exceeds capacity in dimension $d$ at time $t$; $C^d_{jt}$ is the server capacity; and $\mathcal{Y}^d_i \cdot \alpha_{ijt}$ is the actual resource usage of VM $i$ when it is placed on server $j$ and active at time $t$.

We encode \autoref{eq:overutil-jtd} as a MILP using standard techniques (e.g., Big-M and auxiliary variables). We then define a binary variable $\textit{overutil}_{jt}$ that is 1 if the server is overutilized in \emph{any} dimension:
\begin{align}
    &\textit{overutil}_{jt} \ge \textit{overutil}_{jt}^d, \quad \forall j,d,t \notag \\
    &\textit{overutil}_{jt} \le \sum_d \textit{overutil}_{jt}^d, \quad \forall j,t \notag
\end{align}

The overall metric sums over all servers and timesteps:
\begin{align}
    \texttt{Server Overutilization} = \sum_j \sum_t \textit{overutil}_{jt} \notag
\end{align}

\parab{Server utilization.}
We also measure how many servers are active. We define a binary variable $util_{jt}$ that is 1 if at least one VM is assigned to server $j$ at time $t$:
\begin{align}
    util_{jt} \le \sum_i \alpha_{ijt}, \quad \forall j,t \\
    util_{jt} \ge \alpha_{ijt}, \quad \forall i,j,t
\end{align}

We then aggregate this over time:
\begin{align}
    \texttt{Server Utilization} = \sum_j \sum_t util_{jt} \notag
\end{align}
\end{document}